\documentclass[superscriptaddress,twocolumn]{revtex4-2}

\usepackage{amsfonts,graphicx,hyperref,mathtools}

\usepackage{atveryend}
\makeatletter
\let\origcitation\citation
\AtEndDocument{\def\mycites{\@gobble}%
  \def\citation#1{\g@addto@macro\mycites{,#1}\origcitation{#1}}}
\AtVeryEndDocument{\typeout{***^^JCited keys: \mycites^^J***}}
\makeatother

\graphicspath{{figures/}}

\allowdisplaybreaks[1]

\def\guil#1{{\fontencoding{T1}\selectfont\guillemetleft#1\guillemetright}}

\def\vec#1{{\bf #1}}

\begin{document}

\title{Yb delafossites: unique exchange frustration of 4f spin 1/2 moments on a perfect triangular lattice}

\author{B.~Schmidt}
\author{J.~Sichelschmidt}
\author{K.~M.~Ranjith}
\affiliation{Max-Planck-Institut f\"ur chemische Physik fester Stoffe, Dresden, Germany}
\author{Th.~Doert}
\affiliation{Fakult\"at f\"ur Chemie und Lebensmittelchemie, Technische Universit\"at Dresden, Germany}
\author{M.~Baenitz}
\affiliation{Max-Planck-Institut f\"ur chemische Physik fester Stoffe, Dresden, Germany}

\date{Typeset \today}

\begin{abstract}
While the Heisenberg model for magnetic Mott insulators on planar lattice structures is comparatively well understood in the case of transition metal ions, the intrinsic spin-orbit entanglement of 4f magnetic ions on such lattices shows fascinating new physics largely due to corresponding strong anisotropies both in their single-ion and their exchange properties. We show here that the Yb delafossites, containing perfect magnetic Yb$^{3+}$ triangular lattice planes with pseudospin $s=1/2$ at low temperatures, are an ideal platform to study these new phenomena. Competing frustrated interactions may lead to an absence of magnetic order associated to a gapless spin liquid ground state with a huge linear specific heat exceeding that of many heavy fermions, whereas the application of a magnetic field induces anisotropic magnetic order with successive transitions into different long ranged ordered structures. In this comparative study, we discuss our experimental findings in terms of a unified crystal-field and exchange model. We combine electron paramagnetic resonance (EPR) experiments and results from neutron scattering with measurements of the magnetic susceptibility, isothermal magnetization up to full polarization, and specific heat to determine the relevant model parameters. The impact of the crystal field is discussed as well as the symmetry-compatible form of the exchange tensor, and we give explicit expressions for the anisotropic g factor, the temperature dependence of the susceptibility, the exchange-narrowed EPR linewidth and the saturation field.

\end{abstract} 

\maketitle

\section{Introduction}

We would like to discuss our findings and review further available results about an interesting class of compounds -- the Ytterbium Yb$^{3+}$ delafossites. The delafossites in general emerged from the end of the 19th century onwards~\footnote{The original name was given by the French mineralogist Charles Friedel (1832--1899) to the natural mineral CuFeO$_\text2$ in honor of Gabriel Delafosse (1796–1878)~\cite{friedel:73,rogers:13} in 1873.}, the name is structurally motivated and as such has little relationship to the properties of the individual compounds falling into this class -- there are insulators, metals, superconductors, semimetals, quasi two dimensional highly conductive materials, magnetic materials, and more. For a large variety of ground states one may think of it is highly probable that we can find representatives in the delafossite class of minerals.

The delafossites form as $A^{1+}R^{3+}X^{2-}_{2}$, where $A$ is an alkaline metal (Li, Na, K, Rb, Cs) or a monovalent transition metal ion (Pd, Pt or Cu, Ag), $R$ is a trivalent transition metal or rare earth ion which might be magnetic (like Ti, V, Cr, Fe, Ce or Yb) or nonmagnetic (Al, Ga, In, Tl or Co, Rh), and $X$ stands for a chalcogen which is either oxygen, sulfur, or selenium. Most of the compounds form in the rhombohedral  $\alpha$-NaFeO$_{2}$ delafossite structure~\cite{marquardt:06,mackenzie:17} with space group $R\bar3m$.

Until now, studies on 4f delafossites have been very rare. The reason is that the 4f ions are relatively large and are difficult to incorporate into oxygen-based delafossite structures which are widely investigated. For delafossites with sulfur or selenium, however, this is quite possible due to the larger size of voids formed by the the chalcogen and even the growth of sizable single crystals is possible~\cite{schleid:93}. An essential characteristic of the delafossite structure is the presence of triangular planes composed of edge-sharing $R$X$_{6}$ octahedra. In this respect they can serve as model systems for quantum magnetism in a perfect planar triangular lattice. Due to the ideal triangular structure, geometric frustration counteracts or even suppresses magnetic order at low temperatures, eventually supporting spin liquid behavior. Alternatively, at zero temperature, a magnetic order, the so-called 120-degree order, is also predicted~\cite{huse:88,mila:00,white:07,zhou:17}.

Among the magnetic trivalent transition metal ions only Ti$^{3+}$ has an effective spin $s=1/2$ doublet ground state. Unfortunately it turns out that compounds based on Ti such as NaTiO$_2$ show structural instabilities which result in phase transitions and symmetry reductions introducing additional complexity in the interpretation of the results obtained~\cite{hirakawa:85,clarke:98,ramirez:01}. Here the importance of the 4f ions comes into play: Among them, the Kramers ions with an odd number of electrons or holes in the 4f shell like Ce$^{3+}$ or Yb$^{3+}$ have a pronounced doublet ground state due to a low-symmetry crystal electric field (CEF) and can be described with a pseudospin $s=1/2$. In this respect we underscore that Yb delafossites are ideal model systems for the study of spin 1/2 triangular lattices.

\begin{table*}
\caption{Yb delafossites. Lengths of the long ($\lambda$) and short ($\sigma$) edges of the triangles forming the distorted YbCh$_\text6$ octahedra for some MYbCh$_\text2$ compounds with M being a metal and Ch being a chalcogenide. Next column:  tilting angle $\alpha=\cos^{-1}\left(\lambda/(\sqrt3\sigma)\right)$ of the octahedral rectangles with respect to the triangular plane. Last column: Those compounds marked with an asterisk have no magnetic long-range order down to the respective lowest investigation temperature. For all others, this has not (yet) been investigated. Last row: undistorted octahedron.}
\begin{center}
\begin{tabular}{l|lll|l|l|l}
	compound & $\lambda/\textup{\AA}$ & $\sigma/\textup{\AA}$ & $\alpha$ & 
	reference & space group & remark\\
	\hline\hline
	LiYbS$_\text2$&$3.80800$&$3.24599$&$47.35^\circ$&
	\cite{cotter:94,ranjith:20} & $R\bar3m$ & * \\
	NaYbO$_\text2$&$3.3510$&$2.87472$&$47.70^\circ$&
	\cite{hashimoto:03,ranjith:19} & $R\bar3m$ & * \\
	NaYbS$_\text2$&$3.90400$&$3.40822$&$48.60^\circ$&
	\cite{cotter:94,sichelschmidt:19} & $R\bar3m$ & * \\
	NaYbSe$_\text2$&$4.05680$&$3.92235$&$53.33^\circ$&
	\cite{gray:14,ranjith:19a} & $R\bar3m$ & * \\
	KYbO$_\text2$&$3.40010$&$3.01711$&$49.41^\circ$&
	\cite{dong:08} & $R\bar3m$ & * \\
	KYbS$_\text2$&$3.96800$&$3.62569$&$50.81^\circ$&
	\cite{cotter:94,iizuka:20} & $R\bar3m$ & * \\
	KYbSe$_\text2$&$4.11100$&$5.06809$&$62.07^\circ$&
	\cite{gray:14} & $R\bar3m$ \\
	RbYbO$_\text2$&$3.41000$&$3.16297$&$51.51^\circ$&
	\cite{seeger:69} & $R\bar3m$ \\
	RbYbS$_\text2$&$3.99100$&$3.71481$&$51.66^\circ$&
	\cite{bronger:96} & $R\bar3m$ \\
	CsYbS$_\text2$&$4.02200$&$3.70764$&$51.22^\circ$&
	\cite{bronger:93} & $R\bar3m$ \\
	CsYbSe$_\text2$&$4.15390$&$3.88335$&$51.86^\circ$&
	\cite{deng:05,xing:19} & $P6_3/mmc$ & * \\
	\hline
	TlYbS$_\text2$&$3.93500$&$3.62771$&$51.23^\circ$&
	\cite{duczmal:94}& $R\bar3m$ & * \\
	&$3.9454$&$3.726$&$52.31^\circ$&
	\cite{ferreira:20}& $P6_3/mmc$ & * \\
	\hline
	AgYbO$_\text2$&$3.44040$&$2.99493$&$48.45^\circ$&
	\cite{miyasaka:09,sichelschmidt:20} & $R\bar3m$ & *,
	linear Ag coordination\\
	CuYbSe$_\text2$&$4.01670$&$3.93641$&$53.91^\circ$&
	\cite{daszkiewicz:08} & $P\bar3m1$ & tetrahedral Cu coordination\\
	\hline\hline
	undistorted&$1$&$1$&$54.74^\circ$&&(cubic)&\\
\end{tabular}
\end{center}
\label{tbl:ybdelafossites}
\end{table*}
Our research of the available literature returned fourteen Yb delafossite systems, see Table~\ref{tbl:ybdelafossites}. We expect that the number of compounds will increase over time due to the huge interest among the quantum magnetism community. Starting from NaYbS$_2$, we have established the series of NaYbCh$_2$ delafossites as potential quantum spin liquids and will discuss these systems in particular for a comparative analysis, since we consider them to be prototypical~\cite{baenitz:18,ranjith:19,ranjith:19a}.

The most remarkable property of these materials is the absence of magnetic order down to lowest reached temperatures $T=50\,\rm mK$, suggesting that we might have an experimental realization of the theoretically predicted spin-liquid type ground state~\cite{rau:18,zhu:18}. Another striking feature is that upon the application of a magnetic field the nonmagnetic ground state transforms into a long-range ordered antiferromagnetic state. Therefore with the Yb delafossites we are dealing with systems in the vicinity of magnetic order which might be tagged as critical spin liquids. This is the crucial difference to the known putative spin liquid candidates like the triangular lattice organic salts, the kagome type herbertsmithite or the recently discovered hyperkagome iridates which are all far away from magnetic order~\cite{balents:10,savary:17,knolle:19}.

In the following sections, we will try to reconcile our theoretical considerations based on a crystal-field plus nearest-neighbor exchange model with the experimental results. In summary the Yb delafossite compounds are interesting unique systems with an ideal triangular lattice structure which, together with the strong spin-orbit coupling, leads to unusually large spin and exchange anisotropies. In detail, the triangular crystal field splits the spin-orbit entangled Yb$^{3+}$ states into a series of Kramers doublets, the lowest of which in turn results in a complex correlated ground state with a pseudospin $s=1/2$. As a consequence also the magnetic exchange between the Yb$^{3+}$ ions mediated via the orbitals of the surrounding p states of the chalcogen ions becomes complex and bond dependent, similar to the iridate compounds with honeycomb structure.

\section{One Ytterbium ion}

\subsection{Crystallography}
\label{sec:crystal}

Table~\ref{tbl:ybdelafossites} summarizes the known Yb delafossites. Characteristic for the crystal structure of these is a layered composition of sheets of tilted YbCh$_\text6$ octahedra alternating with \guil{filler} planes comprised of alkaline/transition/boron group metal ions. 

\begin{figure}
\includegraphics[width=\columnwidth]{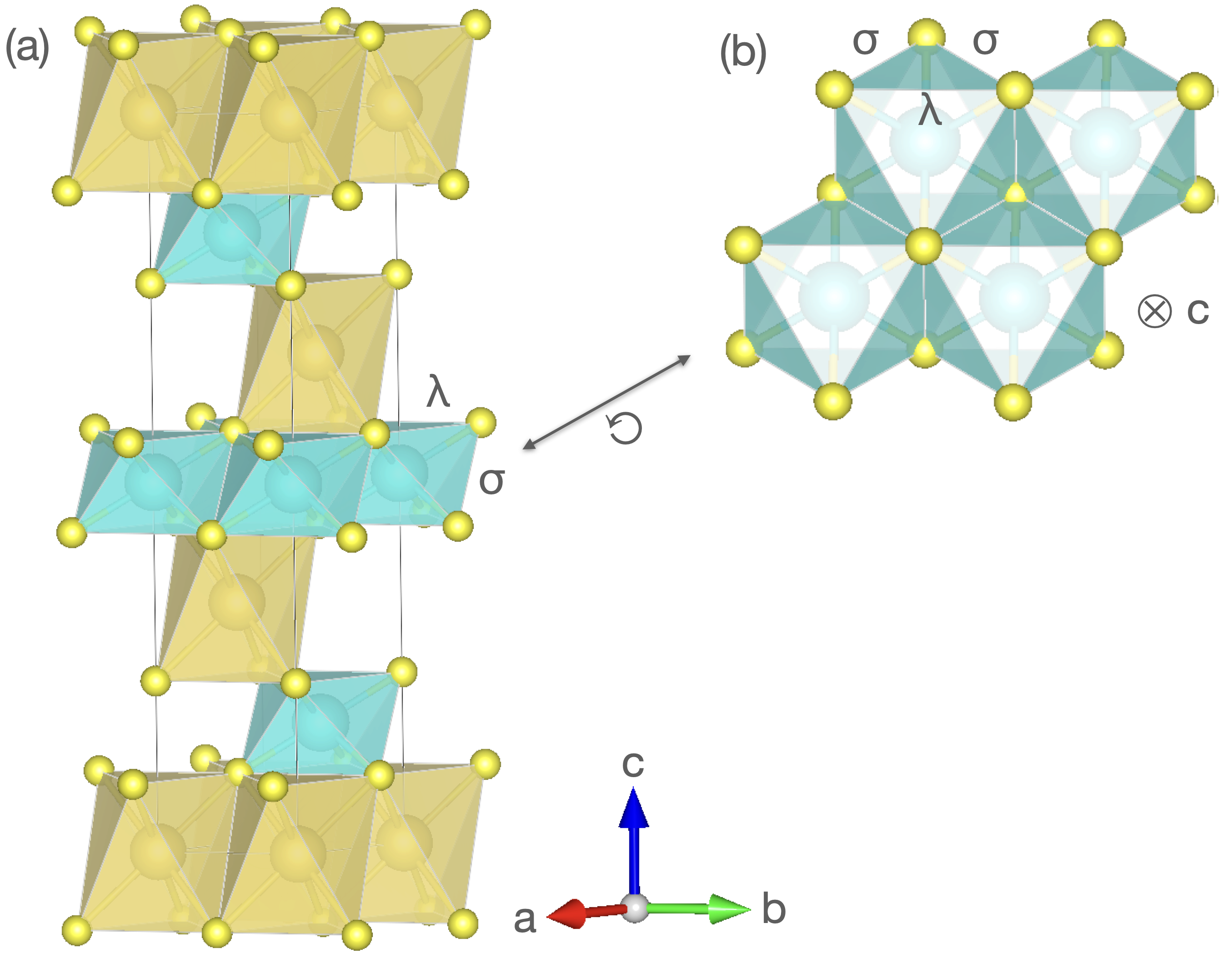}
\caption{Illustration of the crystal structure of NaYbS$_\text2$. The blue polygons represent the distorted YbS$_\text6$ octahedra with Yb$^{3+}$ in the center of their basal planes, the yellow polygons the distorted NaS$_6$ octahedra with the Na$^+$ ions. The S$^{2-}$ ions are printed in yellow. (a) Side view perpendicular to the c direction of the unit cell (thin black lines)  containing three layers of YbS$_\text2$ planes. (b) View from top parallel to the c direction onto the middle layer. The light blue triangles perpendicular to c are equilateral (edge length $\lambda$) and form triangular lattice planes.}
\label{fig:naybs2crystalstructure}
\end{figure}
Fig.~\ref{fig:naybs2crystalstructure} illustrates the crystal structure of NaYbS$_\text2$ as an example. The sulfur octahedra (blue) are tilted such that the Yb$^{3+}$ ions inside form perfect triangular lattice planes perpendicular to the crystallographic c direction. Ideally, an octahedron has four equivalent threefold axes perpendicular to the eight pairwise parallel equilateral triangles forming the surface of it. However, the octahedra of all Yb delafossite compounds are distorted in the same manner: one threefold axis is shortened such that each former octahedron is comprised of two \guil{large} parallel equilateral triangles with edge length $\lambda$ and six \guil{small} isosceles triangles with two edges of length $\sigma$ and one edge of length $\lambda$.  The \guil{large} triangles are those oriented perpendicular to the c direction, forming a perfect triangular lattice. For an ideal octahedron we would have $\sigma=\lambda$, the tilting angle of the octahedral axis with respect to the triangular plane would be $\alpha=\cos^{-1}\left(1/\sqrt3\right)\approx54.74^\circ$. In contrast all delafossites have $\sigma\ne\lambda$, the tilting angle of the octahedral axis then is given by $\alpha=\cos^{-1}\left(\lambda/(\sqrt3\sigma)\right)$, see Table~\ref{tbl:ybdelafossites} for numbers.

The delafossite structure can have two polytypes according to the orientation of the planar layer stacking. The space group of the rhombohedral 3R type delafossites is $R\bar3m$ wheras the hexagonal $2H$ types have a space group of $P6_3/mmc$. The difference between the two polymorphs is the stacking of the planar layers in c direction. Most of the Yb-delafossites belong to the $R\bar3m$ space group (Table~\ref{tbl:ybdelafossites}).  Assigning typical oxidation states, we have Yb$^{3+}$ ions with one hole in the 4f shell and A$^+$ \guil{filler} ions. The latter mostly are alkaline metals, only one Yb delafossite exists with a metal from the Boron group (Tl) and two Yb delafossites have transition metal filler sheets from the Copper group (Ag and Cu).

\subsection{Yb$^\text{3+}$ in a trigonal crystal field}

\begin{figure}
    \centering
    \includegraphics[width=.9\columnwidth]{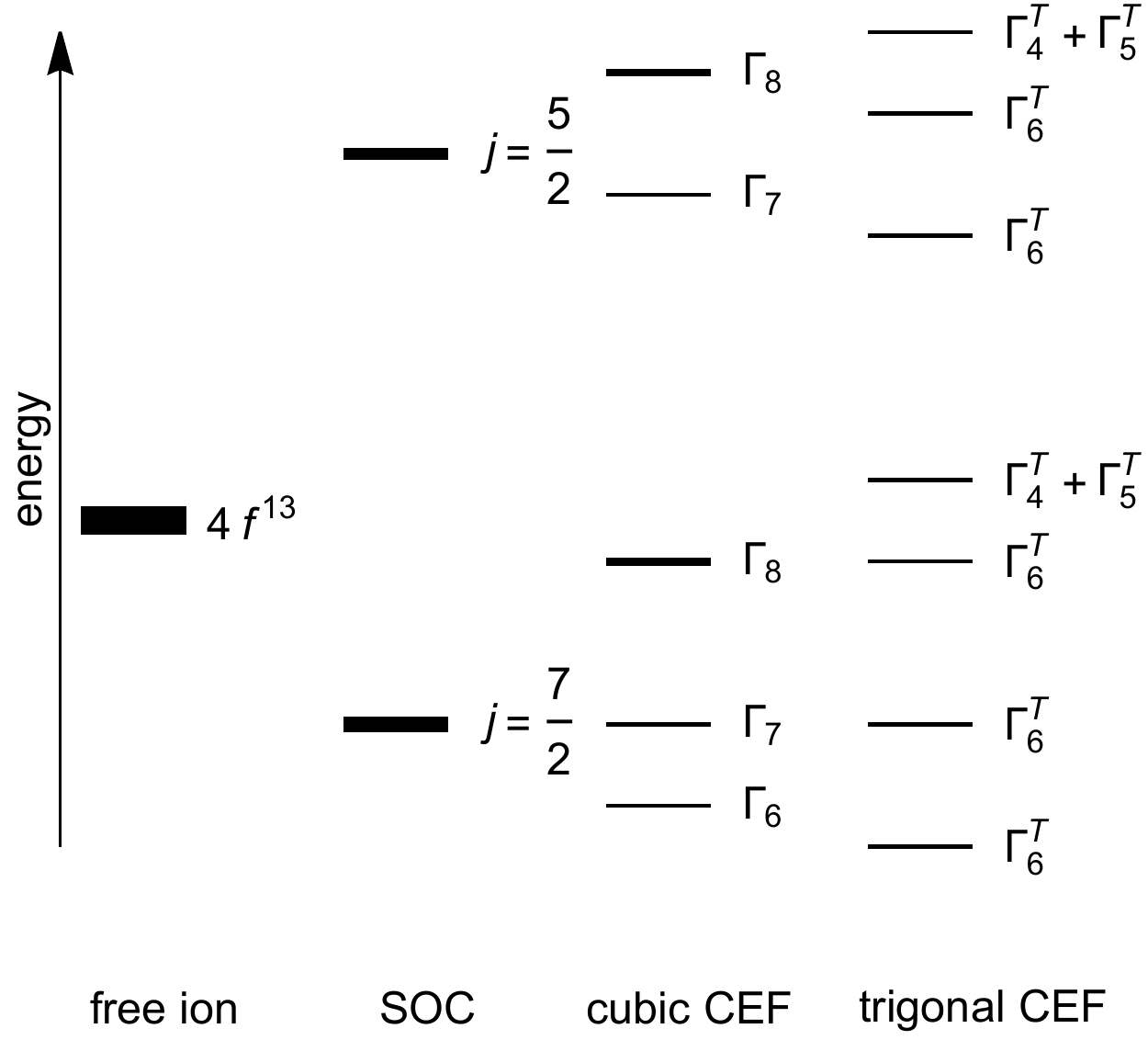}
    \caption{Schematic energy levels of the Yb$^{3+}$ ion (not true to scale). From left to right: free ion, with spin-orbit coupling, in a cubic crystal field, in a trigonal crystal field. The thicknesses of the horizontal lines are proportional to the degeneracies of the respective energy levels, see text.}
    \label{fig:levelscheme}
\end{figure}
Fig.~\ref{fig:levelscheme} schematically shows the energy levels of a single Yb$^{3+}$ ion. The fourteen 4f$^{13}$ states of Yb$^{3+}$ with $\ell=3$, $s=1/2$ are split by the spin-orbit coupling into a $j=\ell+s$ octet and a $j=\ell-s$ sextet. In a perfect octahedral (cubic) environment with ideal tilting angle $\alpha$, the $j=7/2$ states are further split into two doublets $\Gamma_6$ and $\Gamma_7$ and a $\Gamma_8$ quartet, the $j=5/2$ states into a $\Gamma_7$ doublet and a $\Gamma_8$ quartet. Distorting the octahedron along one of its trigonal axes lowers the CEF environment to trigonal, and the local site symmetry of the Yb$^{3+}$ ions is $C_{3\text v}$ with the threefold axis parallel to the c direction. This transforms the formerly cubic states like $\Gamma_6\to\Gamma_6^\text T$, $\Gamma_7\to\Gamma_6^\text T$, and $\Gamma_8\to\Gamma_4^\text T+\Gamma_5^\text T+\Gamma_6^\text T$. We note that although the two representations $\Gamma_4^\text T$ and $\Gamma_5^\text T$ are one-dimensional, due to Kramer's theorem they are complex conjugates and correspond to the same energy. Apart from a constant the Hamiltonian for a single Yb$^{3+}$ ion at an arbitrary lattice site $i$ is then given by
\begin{align}
	{\cal H}_\text{CEF}(i)
	&=
	B_2^0O_2^0({\bf J}_i)+B_4^0O_4^0({\bf J}_i)+B_4^3O_4^3({\bf J}_i)
	\nonumber\\
	&
	+B_6^0O_6^0({\bf J}_i)+B_6^3O_6^3({\bf J}_i)+B_6^6O_6^6({\bf J}_i)
	\label{eqn:cef:htrig}
\end{align}
where $B_n^m$ are crystal-field parameters and $O_n^m({\bf J})$ are Stevens operators, being polynomials of the components of the total-momentum operator~\cite{fulde:79,hutchings:64}. They are reproduced in Appendix~\ref{sec:stevens}.

To gain insight into the structure of the wavefunctions and energy levels, let's for a moment assume the Yb$^{3+}$ ion resides in an ideal octahedron. The local symmetry of the Yb$^{3+}$ ion then is cubic with $O_\text h$ symmetry, and additional relationships between the $B_n^m$ crystal-field parameters apply. Choosing the trigonal axis introduced above as the quantization axis of the hypothetic ideal delafossite, Eq.~(\ref{eqn:cef:htrig}) reduces to~\cite{fulde:79,hutchings:64}
\begin{align}
	{\cal H}_\text{CEF}^{(\text{cubic},3)}
	&=
	B_4^{(3)}\left(O_4^0-20\sqrt2O_4^3\right)
	\nonumber\\
	&\phantom{=}
	+B_6^{(3)}\left(O_6^0+\frac{35\sqrt2}4O_6^3+\frac{77}8O_6^6\right).
	\label{eqn:htri:cubic}
\end{align}
An explicit expression for the matrix of this Hamiltonian for $j=7/2$ in the $\left|j,m\right\rangle$ basis is given in Appendix~\ref{sec:ham:mat}.

Only two independent crystal-field parameters remain. The corresponding wavefunctions of ${\cal H}_\text{CEF}^{(\text{cubic},3)}$ which are grouped in Kramers pairs consisting of time reversed states are given by
\begin{widetext}
\begin{align}
	\Gamma_6
	&:\quad
	\mp\frac13\sqrt{\frac{35}6}\left|\frac72,\pm\frac52\right\rangle
	-\frac13\sqrt{\frac73}\left|\frac72,\mp\frac12\right\rangle
	\pm\frac13\sqrt{\frac56}\left|\frac72,\mp\frac72\right\rangle,
	&\quad
	\left\langle\Gamma_6\right|J_z\left|\Gamma_6\right\rangle
	=\mathop{\rm diag}\left(\pm\frac76\right),
	\nonumber\\
	\Gamma_7
	&:\quad
	\mp\frac13\sqrt{\frac72}\left|\frac72,\pm\frac72\right\rangle
	+\frac{\sqrt5}3\left|\frac72,\pm\frac12\right\rangle
	\pm\frac1{3\sqrt2}\left|\frac72,\mp\frac52\right\rangle,
	&\quad
	\left\langle\Gamma_7\right|J_z\left|\Gamma_7\right\rangle
	=\mathop{\rm diag}\left(\pm\frac32\right),
	\nonumber\\
	\Gamma_8
	&:\quad
	\left\{\begin{aligned}
	&\frac13\sqrt{\frac{14}3}\left|\frac72,\pm\frac72\right\rangle
	\pm\frac13\sqrt{\frac53}\left|\frac72,\pm\frac12\right\rangle
	+\frac23\sqrt{\frac23}\left|\frac72,\mp\frac52\right\rangle\\
	&\left|\frac72,\pm\frac32\right\rangle
	\end{aligned}\right.,
	&\quad
	\left\langle\Gamma_8\right|J_z\left|\Gamma_8\right\rangle
	=
	\mathop{\rm diag}\left(\pm\frac76,\pm\frac32\right).
    \label{eqn:cubicwf}
\end{align}
\end{widetext}
They are independent of the crystal-field parameters and determined by symmetry only. The notation $\mathop{\rm diag}(A)$ means that the matrix representation $\left\langle\Gamma_\alpha\left|J_z\right|\Gamma_\alpha\right\rangle$ of the $J_z$ operator has eigenvalues $A$ with wavefunctions as noted.

\subsection{Ground state of the Yb$^\text{3+}$ ion}
\label{sec:gs}

Lowering the local symmetry of the crystal field from cubic to trigonal by compressing the threefold axis parallel to the c direction splits the $\Gamma_8$ quartet obtained so far into two Kramers doublets and all six crystal field parameters in Eq.~(\ref{eqn:cef:htrig}) become independent. Also the CEF wavefunctions for the four doublets now depend on the crystal-field potential, see Appendix~\ref{sec:ham:mat} for a $j=7/2$ matrix representation of the corresponding Hamiltonian. This Hamiltonian couples only $|j,m\rangle$ states with $\Delta m=\pm3$, such that three of the four resulting Kramers doublets may be written as~\cite{sakai:00,shiba:00}
\begin{align}
	\left|\psi^+\right\rangle
	&
	=
	-\alpha{\rm e}^{{\rm i}\phi_\alpha}\left|\frac72,\frac72\right\rangle
	+\beta\left|\frac72,\frac12\right\rangle
	+\gamma{\rm e}^{-{\rm i}\phi_\gamma}\left|\frac72,-\frac52\right\rangle,
	\nonumber\\
	\left|\psi^-\right\rangle
	&=
	\alpha{\rm e}^{-{\rm i}\phi_\alpha}\left|\frac72,-\frac72\right\rangle
	+\beta\left|\frac72,-\frac12\right\rangle
	-\gamma{\rm e}^{{\rm i}\phi_\gamma}\left|\frac72,\frac52\right\rangle
\label{eqn:cef:doublet}
\end{align}
where $\alpha$, $\beta$, and $\gamma$ are real with $\alpha^2+\beta^2+\gamma^2=1$. We note that time reversal $T$ doesn't change the sign of $\left|\frac72,-\frac12\right\rangle=T\left|\frac72,\frac12\right\rangle$. This is a general feature of the time-reversal operator, giving $T\left|j,\pm\frac12\right\rangle=(-)^{j\pm1/2}\left|j,\mp\frac12\right\rangle$ with $T^2=-1$ for half-integer total momentum $j$.

The fourth doublet is the pure state $\left|\frac72,\pm\frac32\right\rangle$,
adiabatically (as function of the trigonal distortion) connected to the pure state in the $\Gamma_8$ quartet.

One doublet defined in Eqs.~(\ref{eqn:cef:doublet}) will be the ground state. In principle $\left|\frac72,\pm\frac32\right\rangle$ could be the ground-state doublet as well. However, we will see at the end of this section why in our case it is not. With the help of Appendix~\ref{sec:stevens}, the matrix elements of the total momentum operator $\vec J$ within the ground-state doublet can be read off:
\begin{align*}
	\left\langle\psi^\pm\left|J_z\right|\psi^\pm\right\rangle
	&=
	\pm\frac12\left(7\alpha^2+\beta^2-5\gamma^2\right),
	\\
	\left\langle\psi^-\left|J_x\right|\psi^+\right\rangle
	&=
	\sqrt7\alpha\gamma{\rm e}^{{\rm i}(\phi_\alpha-\phi_\gamma)}+2\beta^2
	=\left\langle\psi^+\left|J_x\right|\psi^-\right\rangle^*,
	\\
	\left\langle\psi^-\left|J_y\right|\psi^+\right\rangle
	&=
	{\rm i}\left(\sqrt7\alpha\gamma{\rm e}^{{\rm i}(\phi_\alpha-\phi_\gamma)}+2\beta^2\right)
	\\
	&=\left\langle\psi^+\left|J_y\right|\psi^-\right\rangle^*.
\end{align*}
All other matrix elements vanish. The total momentum operator $\vec J$ transforms like $\vec J=-T\vec JT^{-1}$ under time reversal $T$. With $T=UK$, $U=\exp({\rm i}\pi J_y)$ in our (standard) representation and $K$ the complex conjugation, this requires the matrix elements of $J_z$ and $J_x$ to be real, those of $J_y$ to be purely imaginary, which is equivalent to $\phi_\alpha=\phi_\gamma+2\pi n$, $n\in\mathbb Z$. This indeed allows us to introduce a pseudospin $\vec S$ for the ground-state doublet by mapping
\begin{align*}
    g_jJ_z
    &\to
    g_\parallel S_z,
    \\
    g_jJ_x
    &\to
    g_\perp S_x,
    \\
    g_jJ_y
    &\to
    g_\perp S_y,
    \\
    g_\parallel
    &=
    g_j\left(7\alpha^2+\beta^2-5\gamma^2\right),
    \\
    g_\perp
	&=
	g_j\left(2\sqrt7\alpha\gamma+4\beta^2\right),
	\\
	g_j
	&=
	1+\frac{j(j+1)+s(s+1)-\ell(\ell+1)}{2j(j+1)}
	\nonumber\\
	&=\frac87\approx1.14
	\quad\mbox{(Land\'e factor)}.
\end{align*}
The Zeeman splitting for a single Yb$^{3+}$ ion at site $i$ is then obtained from the Hamiltonian
\begin{align}
    \MoveEqLeft
	{\cal H}_\text{Zeeman}(i)
	=
	-\mu_0g_j\mu_\text B\sum_\alpha J_i^\alpha H_\alpha
	\nonumber\\
	&\to
	-\mu_0\mu_\text B\left[
	g_\parallel S_i^zH_z
	+g_\perp\left(S_i^xH_x+S_i^yH_y\right)
	\right],
	\label{eqn:cef:hzeeman}
\end{align}
where $\mu_0$ is the magnetic permeability constant and $\mu_\text B$ the Bohr magneton.

The above equations give an argument why the pure state $\left|\frac72,\pm\frac32\right\rangle$ cannot be the ground state here: We would have transverse matrix elements $\left\langle\frac72,\pm\frac32\left|J_{x,y}\right|\frac72,\mp\frac32\right\rangle=0$, i.\,e. $g_\perp\equiv0$, and no coupling to a magnetic field applied perpendicular to the threefold axis enforced by trigonal symmetry alone. As we will see, this is in contradiction to experiment. To the best of our knowledge, no Yb compound is known to have the $\left|\frac72,\pm\frac32\right\rangle$ doublet as ground state.

\subsection{Electron paramagnetic resonance}

\begin{figure}
\begin{center}
\includegraphics[width=\columnwidth]{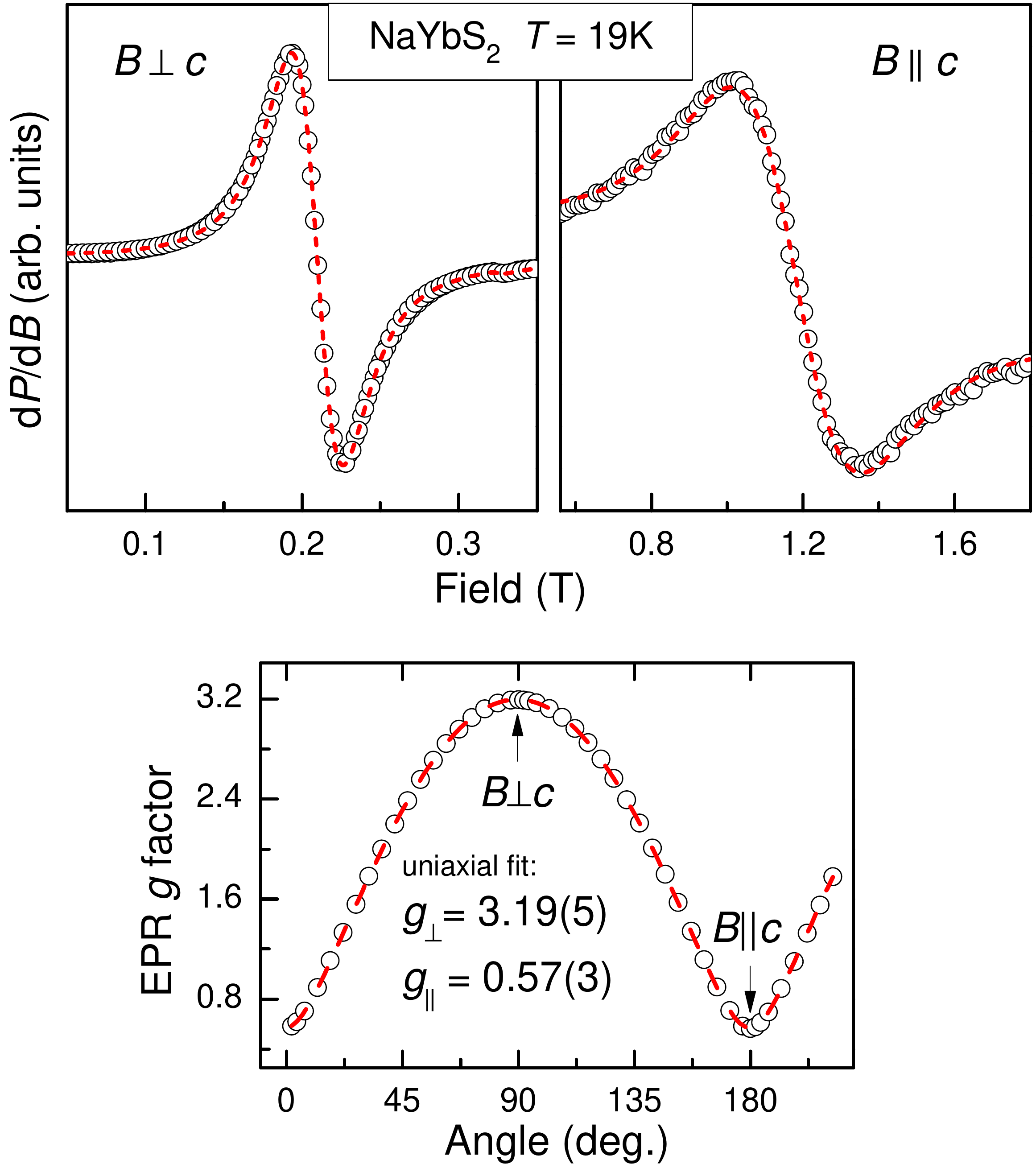}
\end{center}
\caption{
EPR spectra (upper frames) and anisotropy of the EPR $g$ factor (lower frame) of single crystalline NaYbS$_{2}$ at $T=19\,\text K$ and the microwave field $b_\text{mw}\bot\text c$ axis ($\nu=9.4\,\text{GHz}$). The spectra taken at external fields $B\bot\text c$ axis and $B\|\text c$ axis were fitted by a Lorentzian lineshape (dashed lines). The anisotropy of the EPR $g$ factor can be described by $g(\Theta) = \sqrt{g_\|^2\cos^2\Theta + g_\bot^2\sin^2\Theta}$ with $g_\bot=3.19(5)$ and $g_\|=0.57(3)$. The sample was rotated around an axis lying in the basal plane parallel to $b_\text{mw}$. }
\label{ReviewESRFig1}
\end{figure}

We have done thorough electron paramagnetic resonance studies on all three NaYbCh$_\text2$ delafossites~\cite{ranjith:19,sichelschmidt:19,ranjith:19a}. Table~\ref{tbl:values} contains our experimental findings on the g factors. Common remarkable feature is the strong anisotropy between in-plane and out-of-plane gyromagnetism. This is shown exemplarily for NaYbS$_{2}$ in Fig.\ref{ReviewESRFig1}. In a certain temperature range, the resonance intensity $I_\text{EPR}$ is antiproportional to the temperature~\cite{sichelschmidt:20}. Regarding $I_\text{EPR}$ as a measure for the resonant susceptibility $\chi_\text R$, we can correspondingly assign characteristic temperatures $\Theta_{\parallel,\perp}$, see Table~\ref{tbl:values}. We note that these temperatures are not necessarily identical with the Curie-Weiss temperatures obtained from susceptibility measurements (see below).

Towards high temperatures the temperature dependence of the resonance linewidth $\Delta B(T)$ can be understood with a spin-lattice relaxation through the modulation of the ligand field by the lattice vibrations. The basic mechanism behind this relaxation is the spin-orbit coupling which make the electron spins \guil{feel} the ligand field modulation~\cite[p.~60\,ff]{abragam:70}. For the temperature dependence of this relaxation various processes are involved among which an exponential temperature dependence identifies a two-phonon process~\cite{orbach:61}. In this so-called Orbach process the thermal equilibrium of the Zeeman split ground doublet is achieved by a phonon absorption exciting the spin system to an upper state at energy $\Delta E$ and then a phonon emission back to the ground state, the absorption and emission energies differing by the Zeeman energy. This process is determined by the number of phonons at energy $\Delta E$ and yields for $\Delta E\gg k_\text BT$ an approximate temperature dependence $\propto\exp(-\Delta E/k_\text BT)$. Identifying $\Delta E$ with the energy $\Delta E_{12}$ of the first excited crystal-field split state of the Yb$^{3+}$ ion the data analysis of $\Delta B(T)$ gives a rough estimate (within $\pm2.5\,\rm meV$) of $\Delta E_{12}$, values of which are also shown in Table~\ref{tbl:values}.

\subsection{Inelastic neutron scattering}

\begin{table*}
\caption{Characteristic values for the NaYbCh$_\text2$ delafossites, determined by different experimental methods discussed in the text. For NaYbO$_\text2$, only results on powder samples are available which are reproduced in the respective $\parallel$ row of each quantity. An asterisk in one of the Ref. columns denotes this publication.}
\label{tbl:values}
\[
\begin{array}{cc|c|c|c|c|c|c|l}
\text{observable} & \text{quantity} &
\text{NaYbO}_\text2 & \text{Ref.} &
\text{NaYbS}_\text2 & \text{Ref.} &
\text{NaYbSe}_\text2 & \text{Ref.} &
\text{remarks}
\\
\hline\hline
\text{EPR} & \left|g_\parallel\right| &
1.75(3) & \text{\cite{ranjith:19}} &
0.57(3) & \text{\cite{sichelschmidt:19}} &
1.01(1) & \text{\cite{sichelschmidt:20}} &
T=20\,\text K
\\
& \left|g_\perp\right| &
3.28(8) & &
3.19(5) & &
3.13(4) & &
\\
&
\Theta_\parallel &
-9\,\text K & &
-15.2\,\text K & &
-14.3\,\text K & &
\\
& \Theta_\perp &
& &
-14.8\,\text K & &
-14.0\,\text K & &
\\
& \Delta E_{12} &
27\,\text{meV} & &
17\,\text{meV} & &
14\,\text{meV} & &
\\
\hline\hline
\text{INS} & \Delta E_{12} &
34.8\,\text{meV} & \text{\cite{ding:19}} &
12\,\text{meV} & \text{\cite{baenitz:18}} &
15.8\,\text{meV} & \text{\cite{zhang:21}} &
T=5\,\text K
\\
& \Delta E_{13} &
58.5\,\text{meV} & &
23\,\text{meV} & &
24.3\,\text{meV} & &
\\
& \Delta E_{14} &
83.1\,\text{meV} & &
39\,\text{meV} & &
30.5\,\text{meV} & &
\\
\hline\hline
\chi(T) & 
\Theta_\parallel &
-6\,\text K & \text{\cite{ranjith:19}} &
-1.8\,\text K & * &
-3.5\,\text K & \text{\cite{ranjith:19a}} &
\\
& \Theta_\perp &
& &
-11.2\,\text K & &
-7\,\text K & &
\\
& \mu_\parallel &
2.6\,\mu_\text B & &
1.2\,\mu_\text B & &
1.1\,\mu_\text B & &
\\
& \mu_\perp &
& &
2.87\,\mu_\text B & &
2.43\,\mu_\text B & &
\\
\hline
& \hat\Theta_{\parallel,\perp} &
-100\,\text K & * &
-66\,\text K & * &
-66\,\text K & * &
\\
& \mu_\text{eff} &
4.6\,\mu_\text B & &
4.6\,\mu_\text B & &
4.6\,\mu_\text B & &
\\
\hline\hline
M(H) &
\mu_0H_\text{sat}^\parallel &
12\,\text T & \text{\cite{ranjith:19}} &
-- & &
25\,\text T & \text{\cite{ranjith:19a}} &
T=470\,\text{mK}
\\
& \mu_0H_\text{sat}^\perp &
& &
14.7\,\text T & \text{\cite{luther:19}} &
12\,\text T& &
\\
& M_\text{sat}^\parallel &
1.36\,\mu_\text B/\text{Yb}^{3+} & &
0.3\,\mu_\text B/\text{Yb}^{3+} & \text{\cite{baenitz:18}} &
0.49\,\mu_\text B/\text{Yb}^{3+} & &
\\
& M_\text{sat}^\perp &
& &
1.6\,\mu_\text B/\text{Yb}^{3+} & \text{\cite{luther:19}} &
1.5\,\mu_\text B/\text{Yb}^{3+} & &
\\
\hline\hline
\end{array}
\]
\end{table*}
Table~\ref{tbl:values} contains results on inelastic neutron scattering as well. For NaYbO$_2$ and NaYbSe$_2$, three clear maxima in energy scans at temperature $T=5\,\text K$ have been observed~\cite{ding:19,zhang:21},  associated with the energy differences $\Delta E_{1j}$ between the ground state $\left|\psi_1^\pm\right\rangle$ and the excited states $\left|\psi_j^\pm\right\rangle$, $j=2,3,4$. Values are quoted in the table.

\begin{figure}
    \centering
    \includegraphics[width=\columnwidth]{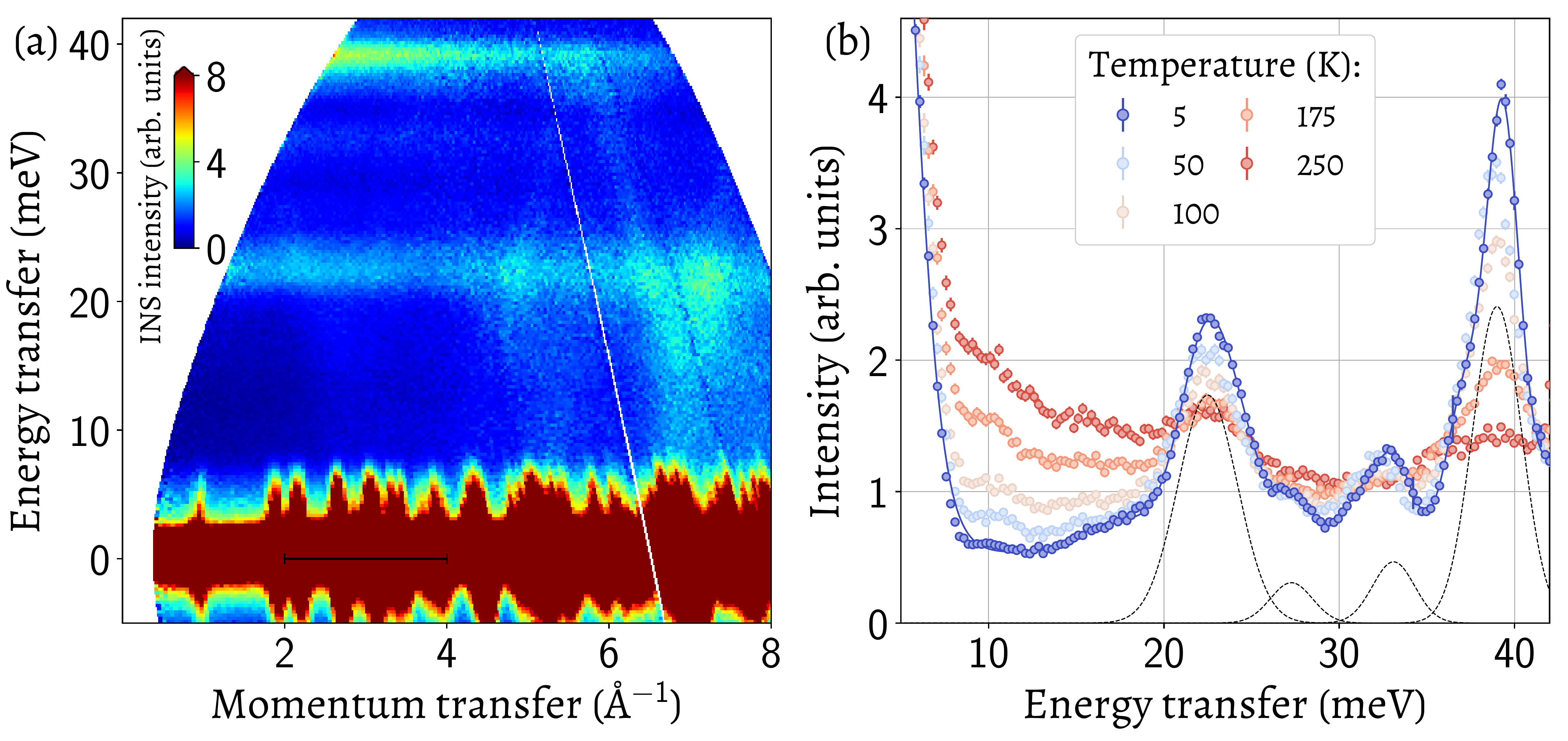}
    \caption{(a) Excitation spectrum of a NaYbS$_\text2$ powder sample measured at $T=5\,\rm K$ with an incident neutron energy $E_\text i=50\,\rm meV$. (b) The spectrum integrated over a momentum transfer range of $2\dots4\,\text\AA^{-1}$ at different temperatures. Dashed line: fit of the $5\,\rm K$ data. Figure taken from Ref.~\cite{baenitz:18}.}
    \label{fig:ins}
\end{figure}
We investigated the excited CEF doublets of NaYbS$_\text2$ in a time-of-flight neutron scattering experiment~\cite{baenitz:18}, see Fig.~\ref{fig:ins} for a summary of the results. At $T=5\,\rm K$ we have found only two clear maxima in the energy-dependent scattering intensity, pointing to excited doublets located $23\,\rm meV$ and $39\,\rm meV$ above the ground state. A third maximum is missing. However at higher temperatures $T\ge50\,\rm K$ additional intensity appears at about $11\,\rm meV$ and $27\,\rm meV$ in the energy scan. If we attribute these features to excitations from the thermally populated first excited doublet to the two higher doublets already observed at low temperatures, we can assume the first excited doublet to be located at an excitation energy of $\Delta E_{12}\approx12\,\rm meV$ above the ground state.

Two qualitative observations can be made here: The higher the g factor anisotropy, (i) the lower the CEF excitation energies, and (ii) the smaller the transition matrix element from the ground state to the first excited doublet is.

As we will see later from analyses of susceptibility data, the typical exchange energy scale of the NaYbCh$_\text2$ delafossites is of the order of a few Kelvin. With the CEF excitations being two orders of magnitude higher, this qualifies the pseudospin description of the ground-state doublet introduced in the prior section also for the minimum exchange model discussed in the following.

\section{Many Ytterbium ions}

\subsection{High-temperature magnetic susceptibility}

The dimensionless uniform magnetic susceptibility $\chi(T)$ of a crystal with volume $V$ is given by the change of the magnetisation $\vec M$ with the magnetic field $\vec B=\mu_0\vec H$ with components
\begin{equation}
	\chi_\alpha=\mu_0\frac{\partial M_\alpha}{\partial B_\alpha}
	=\frac{\mu_0}V\frac{\partial \bar\mu_\alpha}{\partial B_\alpha}
	=-\frac{\mu_0}V\frac{\partial^2F}{\partial B_\alpha^2},
	\quad\alpha=\parallel,\perp
	\label{eqn:chi}
\end{equation}
where $\bar\mu_\alpha=-\partial F/\partial B_\alpha$ is the total magnetic moment in spatial direction $\alpha$ either parallel or perpendicular to the c axis and $F=(1/\beta)\log\cal Z$ the canonical free energy, $1/\beta=k_\text BT$ the inverse temperature and $k_\text B$ the Boltzmann constant. For the molar susceptibility Eq.~(\ref{eqn:chi}) has to be multiplied by $N_\text L/(\nu/V)$ where $N_\text L$ is Avogadro's number and $\nu/V$ the volume density of Yb$^{3+}$ ions.

We link the free energy in the usual way with statistical mechanics through the partition function ${\cal Z}=\exp(-\beta{\cal H})$ where
\begin{equation}
    {\cal H}
    =
    \sum_{i=1}^\nu
    \left[
    {\cal H}_\text{CEF}(i)+{\cal H}_\text{Zeeman}(i)
    +{\cal H}_\text{exc}(i)
    \right]
    \label{eqn:hamall}
\end{equation}
with ${\cal H}_\text{CEF}$ given by Eq.~(\ref{eqn:cef:htrig}), ${\cal H}_\text{Zeeman}$ given by Eq.~(\ref{eqn:cef:hzeeman}), and the exchange Hamiltonian for an arbitrary but fixed site $i$
\begin{equation}
    {\cal H}_\text{exc}(i)
    =
    \frac12\sum_{\langle ij\rangle}
    \sum_{\alpha\beta}
    J_i^\alpha\hat{\cal J}_{ij}^{\alpha\beta}J_j^\beta
    \label{eqn:hexc}
\end{equation}
where the sum is taken over the $z=6$ bonds connecting sites $j$ and site $i$ with an exchange tensor $\hat{\cal J}$ having the respective components $\hat{\cal J}_{ij}^{\alpha\beta}$~\footnote{We use the \guil{hat-notation} to indicate that the corresponding symbol relates to the full angular momentum $J$ with $j=7/2$.}. (See below in Sec.~\ref{sec:pseudospin} for more details on the symmetry-allowed form of the exchange.) A factor $(1/2)$ is included to compensate for double-counting the bonds when executing the sum over the lattice sites in Eq.~(\ref{eqn:hamall}). In the high-temperature limit $\beta\to0$ we can expand $\chi_\alpha$ (Eq.~(\ref{eqn:chi})) in powers of $\beta$. We obtain a Curie-Weiss law
\begin{equation}
	\chi_\alpha
	=
	\frac\nu V\mu_0g_j^2\mu_\text B^2
	\frac{j(j+1)}3\beta\left(
	1+\beta k_\text B\hat\Theta_\alpha
	\right)
	+{\cal O}(\beta^3),
	\label{eqn:chihit}
\end{equation}
with $j=7/2$ and the Curie-Weiss temperatures given by
\begin{align}
	k_\text B\hat\Theta_\parallel
	&=
	-\frac45\left(j-\frac12\right)\left(j+\frac32\right)B_2^0
	-\frac{j(j+1)}3z\hat J_\parallel,
	\nonumber\\
	k_\text B\hat\Theta_\perp
	&=
	+\frac25\left(j-\frac12\right)\left(j+\frac32\right)B_2^0
	-\frac{j(j+1)}3z\hat J_\perp.
	\label{eqn:hatthetacw}
\end{align}
Here $z\hat J_\parallel$ and $z\hat J_\perp$ are the contributions of the exchange tensor parallel and perpendicular to the c direction. Other components of the exchange tensor do not appear in $\chi_\alpha$ up to order $\beta^2$. This result coincides with Ref.~\cite{jensen:91} where a four-parameter crystal field Hamiltonian has been treated. Remarkably, due to the orthogonality and tracelessness of the Stevens operators, only the $B_2^0$ CEF parameter enters the Curie-Weiss temperatures, independent of the form and symmetry of the crystal field otherwise as long as it contains a more-than-2-fold symmetry axis.

\begin{figure}
    \centering
    \includegraphics[width=\columnwidth]{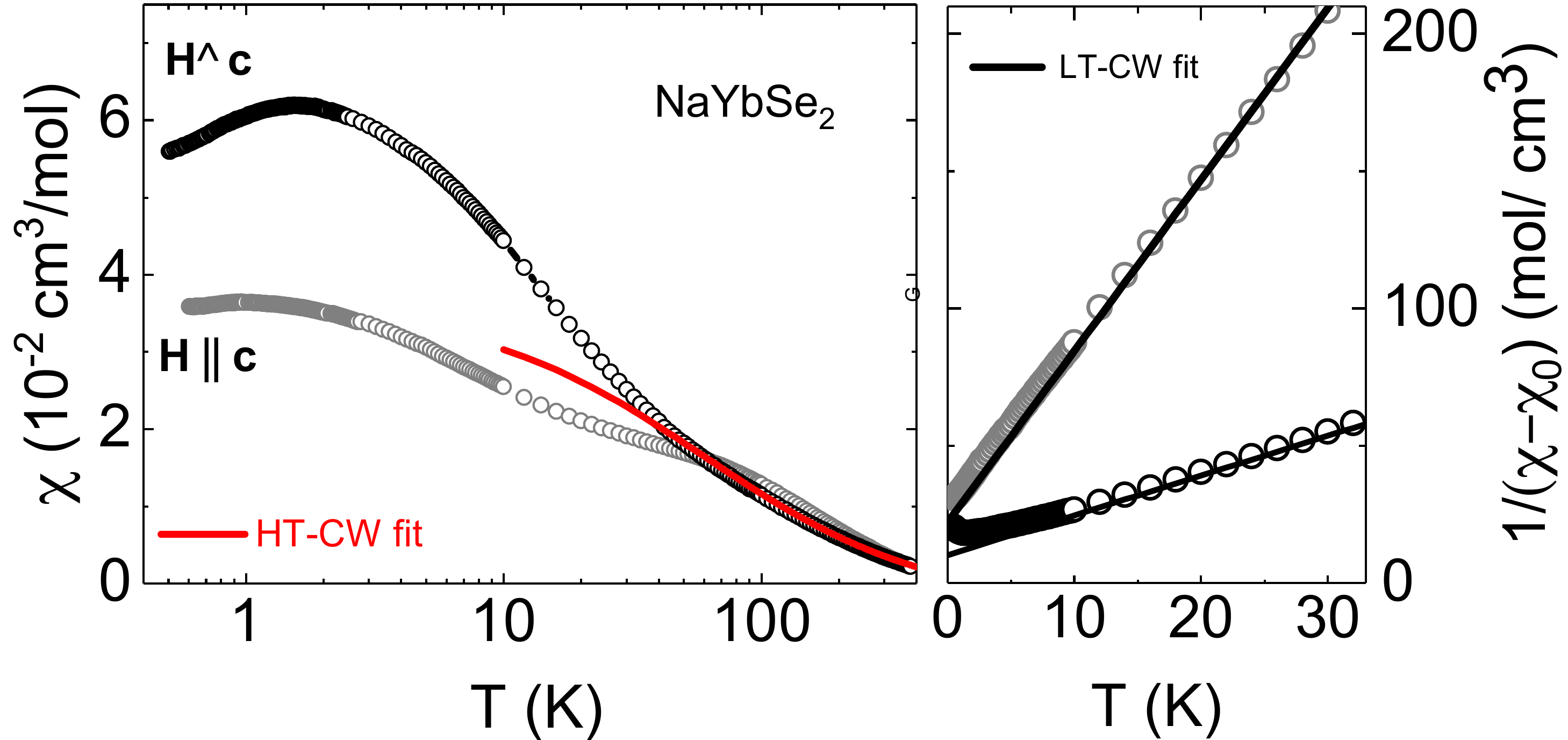}
    \caption{Temperature dependence of the magnetic susceptibility of NaYbSe$_2$ obtained in a field of $1\,\rm T$ for $\chi_{\perp}(T)$ and of $3\,\rm T$ for $\chi_{\parallel}(T)$.}
    \label{fig:chi}
\end{figure}
Figure~\ref{fig:chi} shows the temperature dependence of the susceptibilities $\chi_{\parallel,\perp}(T)$ for NaYbSe$_\text 2$ in the full temperature range $T=0.5\ldots400\,\text K$ accessible to us where we have applied the magnetic field in directions parallel to the c axis (label $H\parallel c$) and perpendicular to it. At temperatures $T\gtrsim120\,\text K$, the inverse $\chi_{\parallel,\perp}^{-1}(T)$ (not shown) show a linear temperature dependence. The solid line in the left plot of Fig.~\ref{fig:chi} denotes a corresponding Curie-Weiss fit to $\chi_\perp(T)$ according to Eq.~(\ref{eqn:chihit}) for $T\ge150\,\text K$. From this fit we obtain $\mu_\text{eff}\approx4.6\,\mu_\text B$, reflecting the full effective moment $\mu_\text{eff}=g_j\mu_\text B\sqrt{j(j+1)}\approx4.54\,\mu_\text B$. We also obtain a Curie-Weiss temperature $\hat\Theta_\perp\approx-66\,\text K$. In the temperature range where the Curie-Weiss law is a good approximation, $\chi_\alpha(T)$ is essentially isotropic with $\hat\Theta_\parallel\approx\hat\Theta_\perp$, independent of the direction of $\bf H$. The same fits for NaYbO$_2$ (powder) and NaYbS$_2$ yield the full moment and isotropic Curie-Weiss temperatures as well. Results are noted in Table~\ref{tbl:values}.

Below $T=80\,\rm K$  the $s=1/2$ pseudospin state emerges.  After subtracting a temperature independent Van-Vleck contribution (obtained from high field magnetization measurements) the susceptibility $\chi(T)$ for $10\,{\rm K}\le T\le40\,\rm K$ can be fitted with a Curie-Weiss law (right-hand plot of Fig.~\ref{fig:chi}) which yields a Curie-Weiss temperature $\Theta_\perp=-7\,\rm K$, and an effective  moment $\mu_\perp=2.43\mu_\text B$ and a Curie-Weiss temperature $\Theta_\parallel=-3.5\,\rm K$ and a moment of  $\mu_\parallel=1.1\mu_\text B$ for fields in the ab plane and in the c direction, respectively. These low-temperature effective moments are consistent with the measured g values from EPR within the $s=1/2$ pseudospin model.

\begin{figure}
    \centering
    \includegraphics[width=\columnwidth]{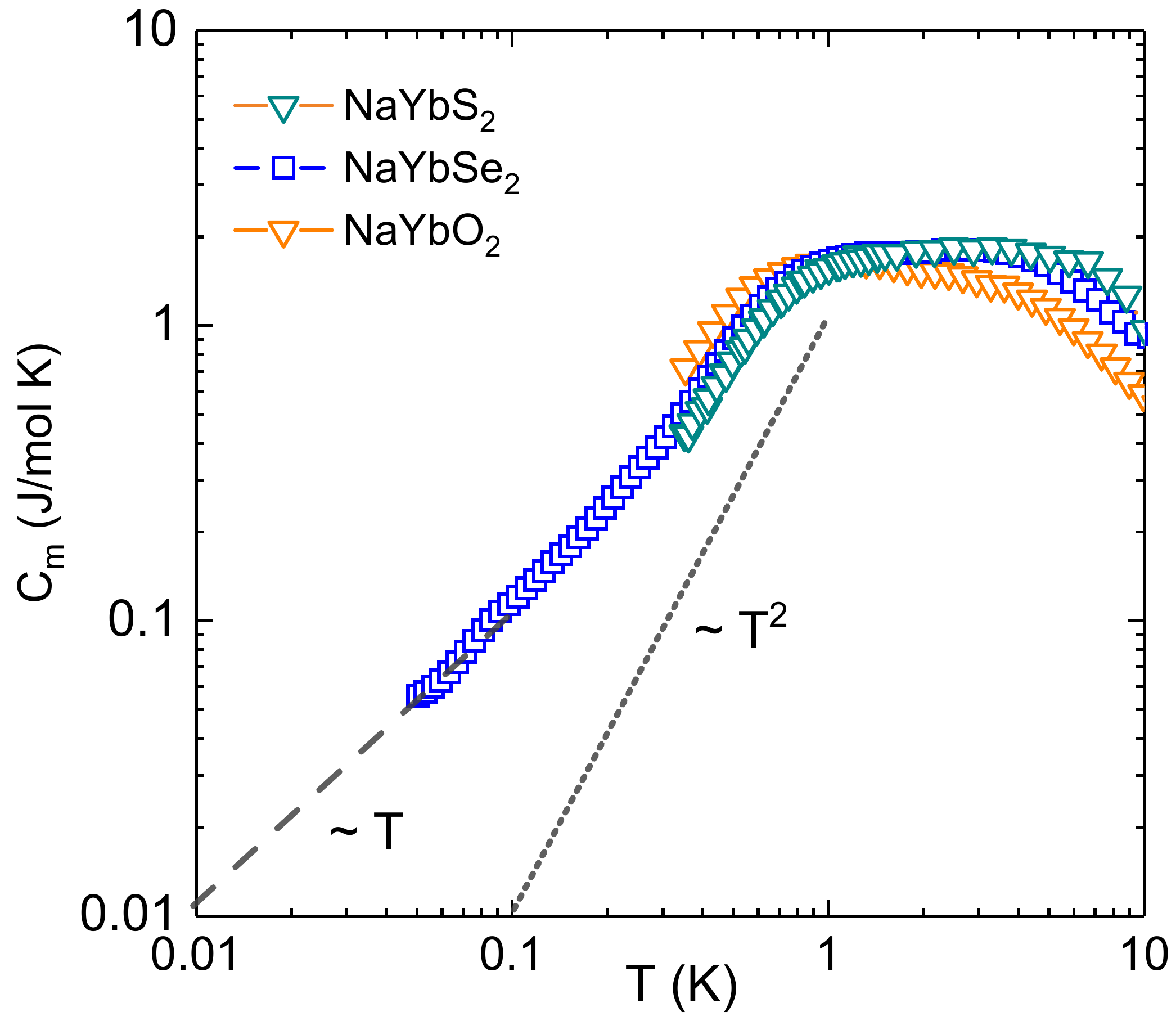}
    \caption{Temperature dependence of the magnetic specific heat of the NaYbCh$_2$ delafossites~\cite{baenitz:18,ranjith:19,ranjith:19a}.}
    \label{fig:cv}
\end{figure}
The pronounced maximum in the susceptibility (Fig.~\ref{fig:chi}) corresponds to the maximum found in the temperature dependence of the magnetic specific heat $c_\text m(T)$ (Fig.~\ref{fig:cv}). Such a maximum is expected in the isotropic triangular lattice~\cite{schmidt:17}. At the maximum, the thermal energy roughly corresponds to the exchange coupling energy of the spin system. In the susceptibility it is clearly visible that for fields in c direction the maximum is at lower temperatures. This is to be expected since the magnetic coupling is smaller in this direction. Fig.~\ref{fig:cv} shows the magnetic specific heat for all three NaYbCh$_2$ compounds. In contrast to the susceptibility, the maximum is broadened here. At temperatures right below the maximum, this specific heat decreases with $T^2$ as expected for two-dimensional magnon-like states, and then a linear dependence $c_\text m(T)=\gamma T$ down to the lowest accessible temperatures with a large residual value $\gamma\approx1\,{\rm J}/\rm molK^2$ is found. This linear temperature dependence is well known for heavy-fermion systems, for example YbRh$_2$Si$_2$~\cite{custers:03} or Yb$_4$As$_3$~\cite{fulde:95,schmidt:96}, yet typical for a gapless spin liquid with fermionic excitations~\cite{knolle:19}.

\subsection{Pseudospin exchange model}
\label{sec:pseudospin}
We restrict ourselves now to the CEF ground-state doublet of each Yb$^{3+}$ ion and use the pseudospin description introduced above. The simplest model Hamiltonian includes nearest-neighbor exchange on the triangular lattice (six neighbors) only. Inspecting one Yb--Yb bond, we see that it contains a twofold rotation axis, a mirror plane in the middle of the bond, and a center of inversion in the middle of the bond. (For an ideal delafossite with undistorted octahedra, we have an additional mirror plane containing the basal plane of an octahedron and a further mirror plane perpendicular to it.) According to Moriya's rules~\cite{moriya:60}, antisymmetric exchange must vanish due to the presence of the inversion center. This leaves us with four independent components of the exchange matrix (three for the ideal case) on any bond $\langle ij\rangle$. We choose our coordinate system such that one bond is parallel to the $x$ direction and the $z$ axis is perpendicular to the triangular-lattice planes. For this bond, we can write
\begin{equation}
	{\cal J}_{ij}
	=
	\begin{pmatrix}
		J_\perp&0&0\\
		0&J_\perp&0\\
		0&0&J_\parallel
	\end{pmatrix}
	+
	\begin{pmatrix}
		J_\Delta&0&0\\
		0&-J_\Delta&J_{yz}\\
		0&J_{yz}&0
	\end{pmatrix},
\end{equation}
whereby we split the exchange matrix into a rotationally invariant part (rotations around the $z$ axis) plus a traceless directional-dependent part. We note that for finite $J_{yz}$ the cartesian coordinate axes $y$ and $z$ are not the main axes of the exchange tensor, rather all three main axis components are different for $\tilde{\cal J}_{ij}=U^{-1}{\cal J}_{ij}U$, $U$ unitary, $\tilde{\cal J}_{ij}$ diagonal. This is a direct consequence of the trigonal distortion (tilting angle $\alpha\ne\cos^{-1}\left(1/\sqrt3\right)$ defined in Sec.~\ref{sec:crystal}) of the YbCh$_6$ octahedra. 

The full pseudospin Hamiltonian with this parametrization, expressed with ladder operators instead of cartesian spin operators then reads
\begin{widetext}
\begin{align}
	{\cal H}
	&=
	\sum_{\left\langle ij\right\rangle}\left\{
	\frac12J_\perp\left(S_i^+S_j^-+S_i^-S_j^+\right)
	+J_\parallel S_i^zS_j^z
	+\frac12J_\Delta\left(
	{\rm e}^{{\rm i}\phi_{ij}}S_i^-S_j^-
	+{\rm e}^{-{\rm i}\phi_{ij}}S_i^+S_j^+
	\right)
	\right.
	\nonumber\\
	&\phantom{=\sum_{\left\langle ij\right\rangle}}\left.
	+\frac1{2\rm i}J_{yz}\left[
	{\rm e}^{{\rm i}\phi_{ij}}\left(S_i^zS_j^++S_i^+S_j^z\right)
	-{\rm e}^{-{\rm i}\phi_{ij}}\left(S_i^zS_j^-+S_i^-S_j^z\right)
	\right]
	\right\}
	+{\cal H}_\text{Zeeman}
	\label{eqn:cef:ham}
\end{align}
\end{widetext}
with ${\cal H}_\text{Zeeman}$ given by Eqs.~(\ref{eqn:cef:hzeeman}) and the direction-dependent phases are
\begin{equation}
	\phi_{ij}
	=
	\left\{
	\begin{array}{rl}
	0,&\vec R_i-\vec R_j=(\pm1,0,0)\\
	\frac{2\pi}3,&
	\vec R_i-\vec R_j=\pm\left(-\frac12,\frac{\sqrt3}2\right)\\
	-\frac{2\pi}3,&
	\vec R_i-\vec R_j=\pm\left(-\frac12,-\frac{\sqrt3}2\right)
	\end{array}
	\right..
\end{equation}
We note that the direction-dependent terms in Eq.(\ref{eqn:cef:ham}) in contrast to the rotationally invariant terms contain at least one spin flip coupling states with $\Delta s_z=1$ or $\Delta s_z=2$.

The three-dimensional ground-state phase diagram of this Hamiltonian has been addressed by several authors~\cite{li:15,li:16,liu:16,luo:17,rau:18,zhu:18,iaconis:18} using slightly differing parameterizations (see Appendix~\ref{sec:params} for examples). Zhu et al.~\cite{zhu:18} have shown that indeed spin-liquid type regions in the ground-state phase diagram of the model~(\ref{eqn:cef:ham}) without magnetic order may exist. However these nonmagnetic regions are comparatively small and require a special choice of the exchange constants. Much more common are magnetically ordered states like stripe phases and the 120-degree pattern known for the isotropic case. Nonetheless of the 14 compounds listed in Table~\ref{tbl:ybdelafossites} at least nine do {\em not\/} show any signature of long-range magnetic order down to the lowest measurement temperature, typically either $0.4\,\rm K$ or $2\,\rm K$. Given the narrowness of the nonmagnetic regions of the model~(\ref{eqn:cef:ham}), it would be  surprising if all compounds can be described with exchange parameters leading to ground states in those regions.

We are faced with a number of possible issues: First, crystallographic peculiarities. A symmetry-allowed buckling of the YbCh$_6$ octahedral planes possibly introduces additional exchange frustration not contained in the model above. Another problem are stacking faults: One unit cell contains three crystallographically equivalent Yb$^{3+}$ ions, each being member of a different YbCh$_6$ distorted-octahedra plane. Adjacent planes are stacked in an A--B--C like fashion where the projections along the c direction of the positions of the Yb$^{3+}$ ions of the \guil{next} layer fall in the middle points of the triangular lattice formed by the \guil{current} layer. This stacking might be distorted. Compatible with these two effects is a sample dependence of the EPR data for NaYbS$_2$ we have observed: The measurements were made for single crystals from two different batches~\cite{sichelschmidt:19}. While the resonance field is essentially identical in both measurements, a pronounced difference has been observed in the width of the EPR resonance. For the smaller crystal, a description of the latter using two Lorentzian lines has been necessary, indicating that roughly half of all spin probes has a larger linewidth than the other half. However, further investigations have to be undertaken to clarify this.

Second, a trivial reason for no magnetic order would be a frustration-induced extremely low ordering temperature $T_\text N$. According to Ref.~\cite{schmidt:17a}, frustration ratios $f=\left|\Theta_{\text{CW}}\right|/T_\text N$~\cite{obradors:88,ramirez:94} of the order of $10$ can be achieved already for an isotropic Heisenberg exchange model on the triangular lattice with $J_\perp=J_\parallel$ and $J_\Delta=J_{yz}=0$ and a small interplane exchange coupling $J_\text{inter}$. With typical Curie-Weiss temperatures of a few Kelvin, it might be that $T_\text N$ falls into the few $100\,\text{mK}$ range, however this requires extremely small $J_\text{inter}={\cal O}(10^{-4}J_{\parallel,\perp})$.

Third, we cannot exclude off-stoichiometric Yb$^{3+}$ ions, and also Na vacancies might be present. Both effects introduce an unknown amount of disorder in the exchange constants, suppressing the magnetic order~\cite{dey:20}.

Fourth, the perfect threefold symmetry of the magnetic sublattice, including the inversion center in the middle of an Yb--Yb bond, might be distorted, introducing changes in bond angles and additional nonzero elements in the exchange matrix. Synchrotron data taken at low temperatures would clarify that.

Fifth, related to the comparatively large local moment of the Yb$^{3+}$ ions in the ab plane (see Table~\ref{tbl:values}), the impact of the long-ranged dipole-dipole interaction might be a further reason for a suppression of $T_\text N$ below our accessible temperature range~\cite{wu:19}.

Finally it might well be that a pure nearest-neighbor exchange model is not sufficient. Additional competing exchange between further neighbors might as well lead to a suppression of a magnetically ordered ground state~\cite{maksimov:19,sedrakyan:20}. Nevertheless we continue using this model for reasons which will become clear later.

\subsection{Electron paramagnetic resonance}

The exchange-narrowed linewidth of the EPR resonance in general is given by
\begin{equation}
	\mu_0\Delta H(\theta)
	=
	\frac{\rm const.}{\mu_\text Bg(\theta)}\,
	M_2\sqrt{\frac{M_2}{M_4}},
	\label{eqn:cef:lw}
\end{equation}
where ${\rm const.}=\pi/\sqrt3$ for a cutoff Lorentzian lineshape, and ${\rm const.}=\sqrt{2\pi}$ for a Lorentzian $\times$ Gaussian lineshape~\cite{castner:71}. $\theta$ is the angle of the applied field $H$ relative to a given crystallographic direction, for example the c axis. $M_2$ and $M_4$ denote the second and fourth moment of the EPR lineshape function, respectively, given by~\cite{zorko:08}
\begin{align}
	M_2
	&=
	\frac{\left\langle
	\left[{\cal H},S_\text{total}^+\right]
	\left[S_\text{total}^-,{\cal H}\right]
	\right\rangle}{\left\langle
	S_\text{total}^+S_\text{total}^-
	\right\rangle},
	\label{eqn:cef:mtwo}
	\\
	M_4
	&=
	\frac{\left\langle
	\left[{\cal H},\left[{\cal H},S_\text{total}^+\right]\right]
	\left[\left[S_\text{total}^-,{\cal H}\right],{\cal H}\right]
	\right\rangle}{\left\langle
	S_\text{total}^+S_\text{total}^-
	\right\rangle},
	\label{eqn:cef:mfour}
	\\
	S_\text{total}^\pm
	&=
	\sum_iS_i^\pm.
\end{align}
Here the $z$ axis not necessarily corresponds to the crystallographic threefold c axis but rather is defined by the direction of the applied field. Appendix~\ref{sec:moment} contains more details on the calculation of $M_{2,4}$.

For a Gaussian lineshape, all odd moments vanish, the higher even moments all factorize into powers of the second moment, and we have
\begin{align}
	M_4^\text{Gauss}
	&=
	3\left(M_2^\text{Gauss}\right)^2,
	\\
	\mu_0\Delta H_\text{Gauss}(\theta)
	&=
	\frac{\text{const.}}{\mu_\text Bg(\theta)}
	\sqrt{\frac{M_2^\text{Gauss}}3}.
\end{align}
Using the Hamiltonian~(\ref{eqn:cef:ham}), we obtain
\begin{align}
	\lim_{T\to\infty}M_2^\parallel
	&=
	\frac34\left[
	2\left(J_\parallel-J_\perp\right)^2+2J_\Delta^2+5J_{yz}^2
	\right],
	\label{eqn:cef:m2para}
	\\
	\lim_{T\to\infty}M_2^\perp
	&=
	\frac38\left[
	2\left(J_\parallel-J_\perp\right)^2+10J_\Delta^2+7J_{yz}^2
	\right],
	\label{eqn:cef:m2perp}
\end{align}
where the symbols $\parallel$, $\perp$ denote the direction of the applied magnetic field $\mu_0H$ relative to the crystallographic c direction. We note that for a fully isotropic exchange Hamiltonian the expressions above vanish identically. In this case, a finite linewidth is due to dipole-dipole interaction which by its very nature is anisotropic.

Furthermore we read off $M_2^\parallel=2M_2^\perp$ for a rotationally invariant exchange ($J_\Delta=J_{xy}=0$) which can be understood in the following way: Thermal spin fluctuations are energetically favorable perpendicular to the field $\mu_0H$, maximizing the energy gain due to the Zeeman energy by leaving the component of the total moment (anti-) aligned to the field unchanged. Thermal fluctuations out of the crystallographic ab plane are suppressed because they would break the threefold rotational symmetry around the c axis. For a field parallel to c, we therefore have two possible fluctuation directions, for a field in the ab plane only one fluctuation direction remains.

More algebra has to be done to calculate the fourth moment. We eventually obtain
\begin{align}
	\lim_{T\to\infty}M_4^\parallel
	&=
	3\left(J_\perp-J_\parallel\right)^2
	\left(7J_\parallel^2-6J_\parallel J_\perp+11J_\perp^2\right)
	\nonumber\\
	&\phantom={}+
	\frac32\left[
	22J_\Delta^4+71J_\Delta^2J_{yz}^2+76J_{yz}^4
	\right.
	\nonumber\\
	&\phantom{=\frac32}+
	J_\Delta^2\left(
	44J_\perp^2-56J_\perp J_\parallel+24J_\parallel^2
	\right)
	\nonumber\\
	&\phantom{=\frac32}+
	\left.
	J_{yz}^2\left(
	71J_\perp^2-104J_\perp J_\parallel+63J_\parallel^2
	\right)
	\right],
	\label{eqn:cef:m4para}
	\\
	\lim_{T\to\infty}M_4^\perp
	&=
	\frac32\left(J_\perp-J_\parallel\right)^2
	\left(7J_\parallel^2-6J_\parallel J_\perp+11J_\perp^2\right)
	\nonumber\\
	&\phantom={}+
	\frac34\left[
	182J_\Delta^4+193J_\Delta^2J_{yz}^2+92J_{yz}^4
	\right.
	\nonumber\\
	&\phantom{=\frac34}+
	J_\Delta^2\left(
	60J_\perp^2-56J_\perp J_\parallel+56J_\parallel^2
	\right)
	\nonumber\\
	&\phantom{=\frac34}+
	\left.
	J_{yz}^2\left(
	81J_\perp^2-104J_\perp J_\parallel+65J_\parallel^2
	\right)
	\right].
	\label{eqn:cef:m4perp}
\end{align}
Similar to $M_2$, also $M_4$ vanishes for a fully isotropic exchange Hamiltonian, and $M_4^\parallel=2M_4^\perp$ for a rotationally invariant exchange.

We assume $J_\Delta,J_{yz}\ll J_\perp,J_\parallel$. Taking into account only finite exchange constants $J_\perp$ and $J_\parallel$, we get for the high-temperature EPR linewidth, Eq.~(\ref{eqn:cef:lw})
\begin{align}
	\lim_{T\to\infty}\mu_0\Delta H_\parallel
	&=
	\frac{\rm const.}{\mu_\text Bg_\parallel}\frac3{2\sqrt2}
	\frac{\left(J_\perp-J_\parallel\right)^2}
	{\sqrt{7J_\parallel^2-6J_\parallel J_\perp+11J_\perp^2}},
	\label{eqn:lwpara}
	\\
	\lim_{T\to\infty}\mu_0\Delta H_\perp
	&=
	\frac{\rm const.}{\mu_\text Bg_\perp}\frac3{4\sqrt2}
	\frac{\left(J_\perp-J_\parallel\right)^2}
	{\sqrt{7J_\parallel^2-6J_\parallel J_\perp+11J_\perp^2}}
	\label{eqn:lwperp}
\end{align}
for field parallel and perpendicular to the c axis. An estimate of $\lim_{T\to\infty}\Delta H_{\parallel,\perp}$ from our linewidth data is difficult to obtain, because in this limit, phonon-dominated relaxation mechanisms like the Orbach process discussed above might dominate. In particular we have $\lim_{T\to\infty}\Delta H_\parallel/\Delta H_\perp=2g_\perp/g_\parallel$. This relation is roughly consistent with the experimental EPR data. Estimating $\lim_{T\to\infty}\Delta H$ with the smallest value the linewidth reaches in its temperature dependence we obtain for NaYbCh$_{2}$, $\text{Ch}=\text O$, S, Se:  $\lim_{T\to\infty}\Delta H_\parallel/\Delta H_\perp= 1.7 / 10 / 4.6$. From the EPR data in Table~\ref{tbl:values}, we obtain $2g_\perp/g_\parallel= 3.7 / 11.1 / 6.2$.

\subsection{Magnetization and susceptibility}

In order to learn more about the size of the exchange constants, we have measured the temperature-dependent uniform magnetic susceptibility $\chi(T)$ of the pseudospins at sufficiently low temperatures $T<30\,\rm K$ and the magnetization $M(H)$ in applied magnetic fields $B=\mu_0 H$ up to $B=30\,\rm T$. The dimensionless magnetic susceptibility is given by Eq.~(\ref{eqn:chi}). Here we expand the thermal traces for the pseudospin Hamiltonian Eq.~(\ref{eqn:cef:ham}) in the low-temperature limit $\beta\to0$ and obtain
\begin{equation}
	\chi_\alpha
	=
	\frac\nu V\mu_0g_\alpha^2\mu_\text B^2\frac{s(s+1)}3\beta
	\left(1+\beta k_\text B\Theta_\alpha\right)
	+{\cal O}\left(\beta^3\right)
\end{equation}
where the Curie-Weiss temperatures $\Theta_{\parallel,\perp}$ are given by
\begin{align}
	k_\text B\Theta_\alpha
	&=
	-\frac{s(s+1)}3\sum_{n=1}^6{\cal J}_{i,i+n}^\alpha,
	\nonumber\\
	k_\text B\Theta_\parallel
	&=
	-\frac32J_\parallel,
	\quad
	k_\text B\Theta_\perp
	=
	-\frac32J_\perp
	\label{eqn:cef:thetacw}
\end{align}
with $s=1/2$ for a field applied parallel and perpendicular to the crystallographic c direction. Here ${\cal J}_{i,i+n}^\alpha$ denotes the component of the exchange energy between site $i$ and its $n$th neighbor along the field direction $\alpha$. We note that $\Theta_{\parallel,\perp}$ do {\em not\/} depend on the direction-dependent terms in the Hamiltonian~(\ref{eqn:cef:ham}) -- the exchange constants $J_\Delta$ and $J_{yz}$ only appear in higher orders of the expansion (compare also Ref.~\cite{li:15}). 

\begin{table}
\caption{Characteristic Hamiltonian parameters for the NaYbCh$_\text2$ delafossites, calculated from the experimental values in Table~\ref{tbl:values} as discussed in the text.}
\label{tbl:calc}
\[
\begin{array}{cc|c|c|c}
\text{observable} & \text{parameter} &
\text{NaYbO}_\text2 &
\text{NaYbS}_\text2 &
\text{NaYbSe}_\text2
\\
\hline\hline
\chi(T) & g_\parallel &
3.0 &
1.4 &
1.3
\\
& g_\perp &
&
3.3 &
2.8
\\
& J_\parallel &
0.34\,\text{meV} &
0.10\,\text{meV} &
0.20\,\text{meV}
\\
& J_\parallel/k_\text B &
4\,\text K &
1.2\,\text K &
2.3\,\text K
\\
& J_\perp &
&
0.64\,\text{meV} &
0.40\,\text{meV}
\\
& J_\perp/k_\text B &
&
7.5\,\text K &
4.7\,\text K
\\
\hline\hline
M(H) & g_\parallel &
2.7 &
0.6 &
1.0
\\
& g_\perp &
&
3.2 &
3.0
\\
& J_\parallel &
0.42\,\text{meV} &
-- &
0.25\,\text{meV}
\\
& J_\parallel/k_\text B &
4.9\,\text K &
-- &
2.9\,\text K
\\
& J_\perp &
&
0.61\,\text{meV} &
0.46\,\text{meV}
\\
& J_\perp/k_\text B &
&
7.0\,\text K &
5.4\,\text K
\\
\hline\hline
\end{array}
\]
\end{table}
Table~\ref{tbl:values} holds our findings from susceptibility measurements~\cite{ranjith:19,ranjith:19a}. It contains the measured Curie-Weiss temperatures $\Theta_{\parallel,\perp}$ and the effective moments $\mu_{\parallel,\perp}$ obtained from a Curie-Weiss fit to $\chi(T)$ at $T<30\,\text K$. For NaYbO$_\text2$ only powder samples were available, the corresponding averaged values are listed in the respective $x_\parallel$ rows. We can calculate the exchange constants for the parameters of the pseudospin Hamiltonian in  Eq.~(\ref{eqn:cef:ham}) from the Curie-Weiss temperatures given by Eqs.~(\ref{eqn:cef:thetacw}), and with the effective moments given by $\mu_{\parallel,\perp}=g_{\parallel,\perp}\sqrt{s(s+1)}$, we can determine the effective g factors. Results are listed in Table~\ref{tbl:calc}.

Table~\ref{tbl:values} also holds the values for the effective moment $\mu_\text{eff}$ and the Curie-Weiss temperatures $\hat\Theta_{\parallel,\perp}$ for $T>150\,\text K$, in the high-temperature limit the latter are given by Eqs.~(\ref{eqn:hatthetacw}). From the derivation of the ground state pseudospin in Sec~\ref{sec:gs} we have a relationship
\begin{equation}
    J_\alpha=\left(\frac{g_\alpha}{g_j}\right)^2
    \hat J_\alpha,
    \quad
    \alpha=\parallel,\perp
\end{equation}
with the exchange constants introduced in Eq.~(\ref{eqn:hexc}) for the full angular momentum. Together with the EPR g factors given in Table~\ref{tbl:values}, this allows us to roughly estimate the $B_2^0$ CEF parameter of NaYbS$_\text 2$ and NaYbSe$_\text 2$ to $B_2^0\approx-1\ldots{-0.5}\,\text{meV}$.

\subsection{Saturation field}

\begin{figure*}
    \centering
    \includegraphics[width=.3\textwidth]{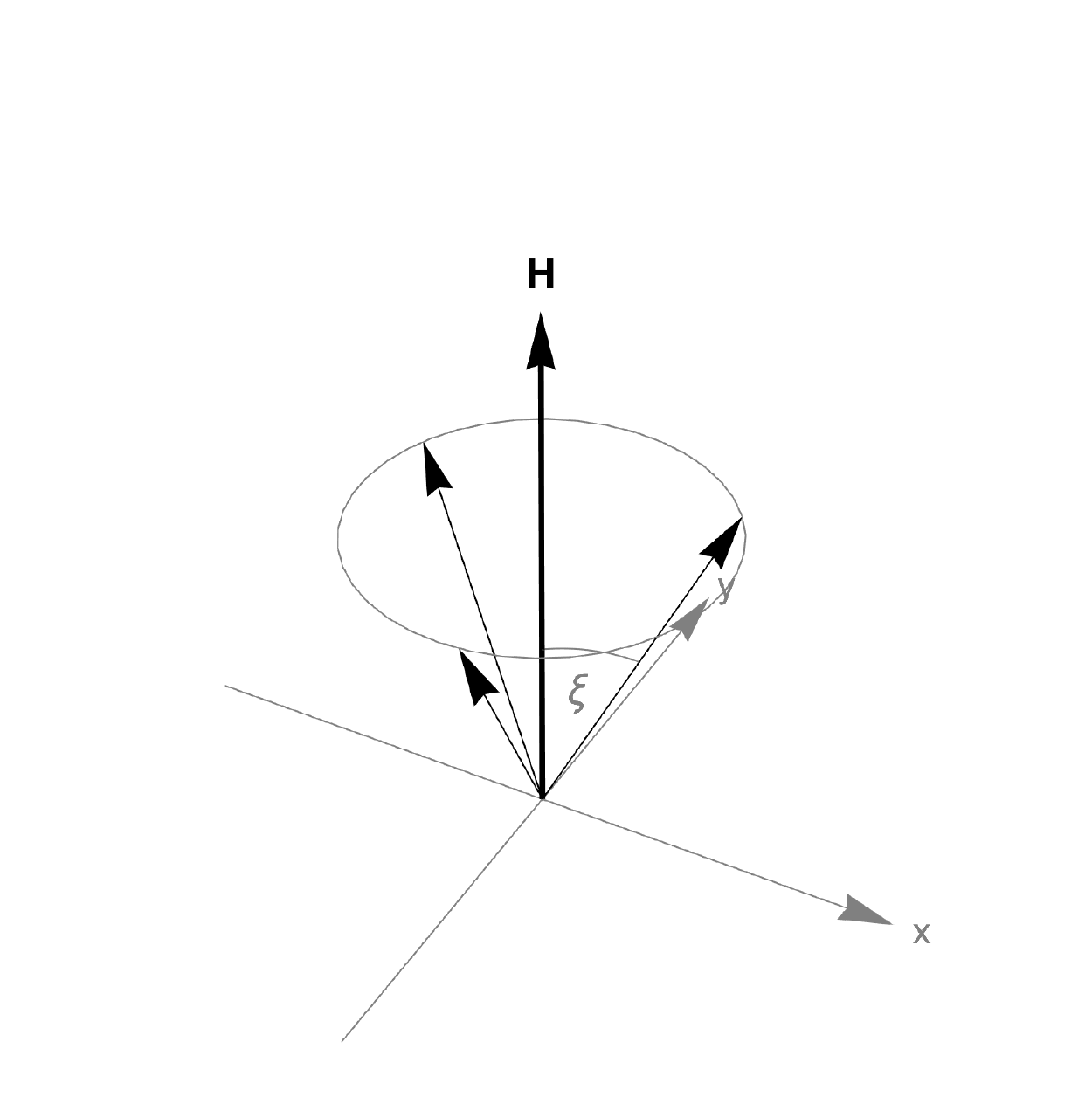}
    \includegraphics[width=.3\textwidth]{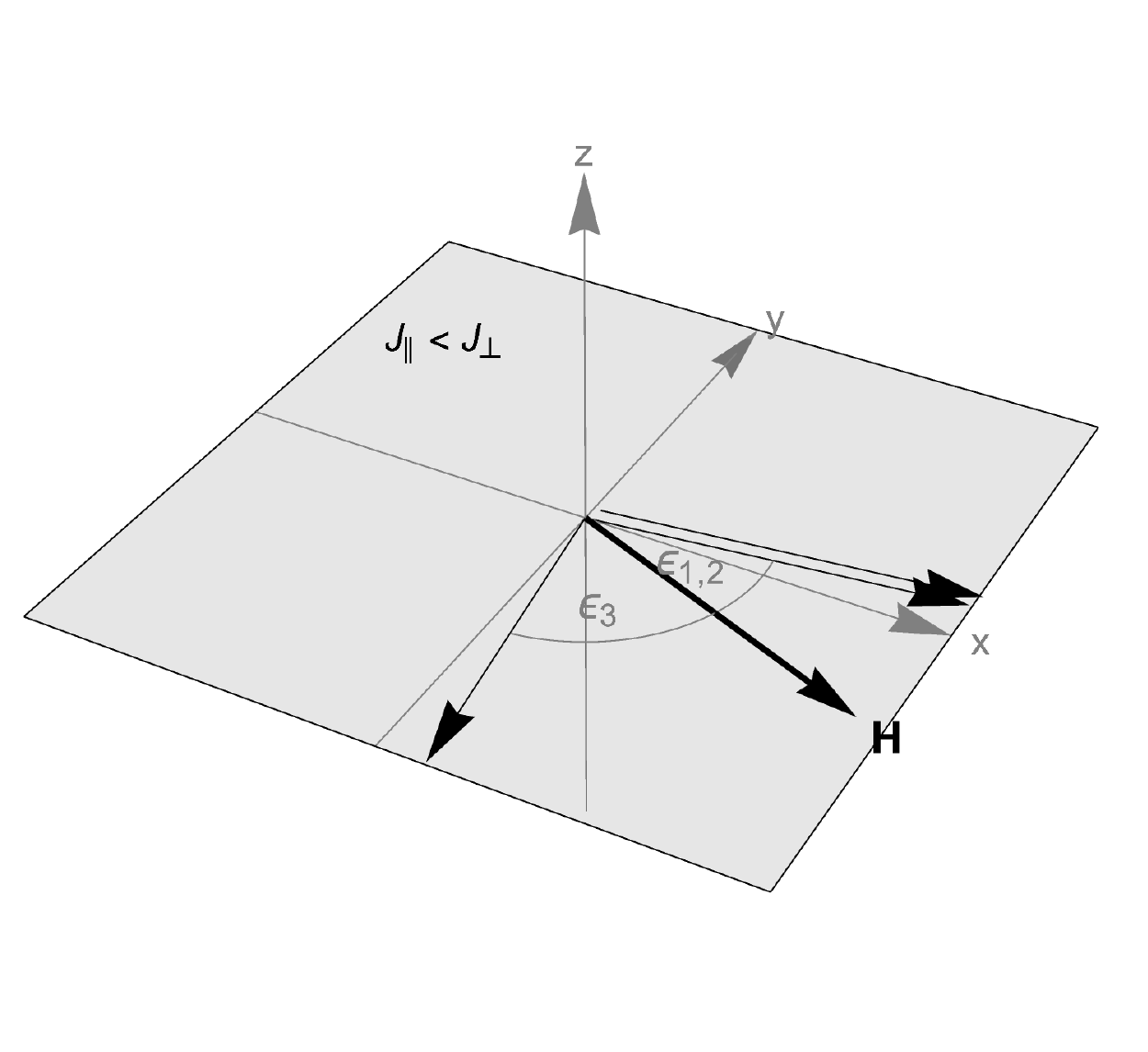}
    \includegraphics[width=.3\textwidth]{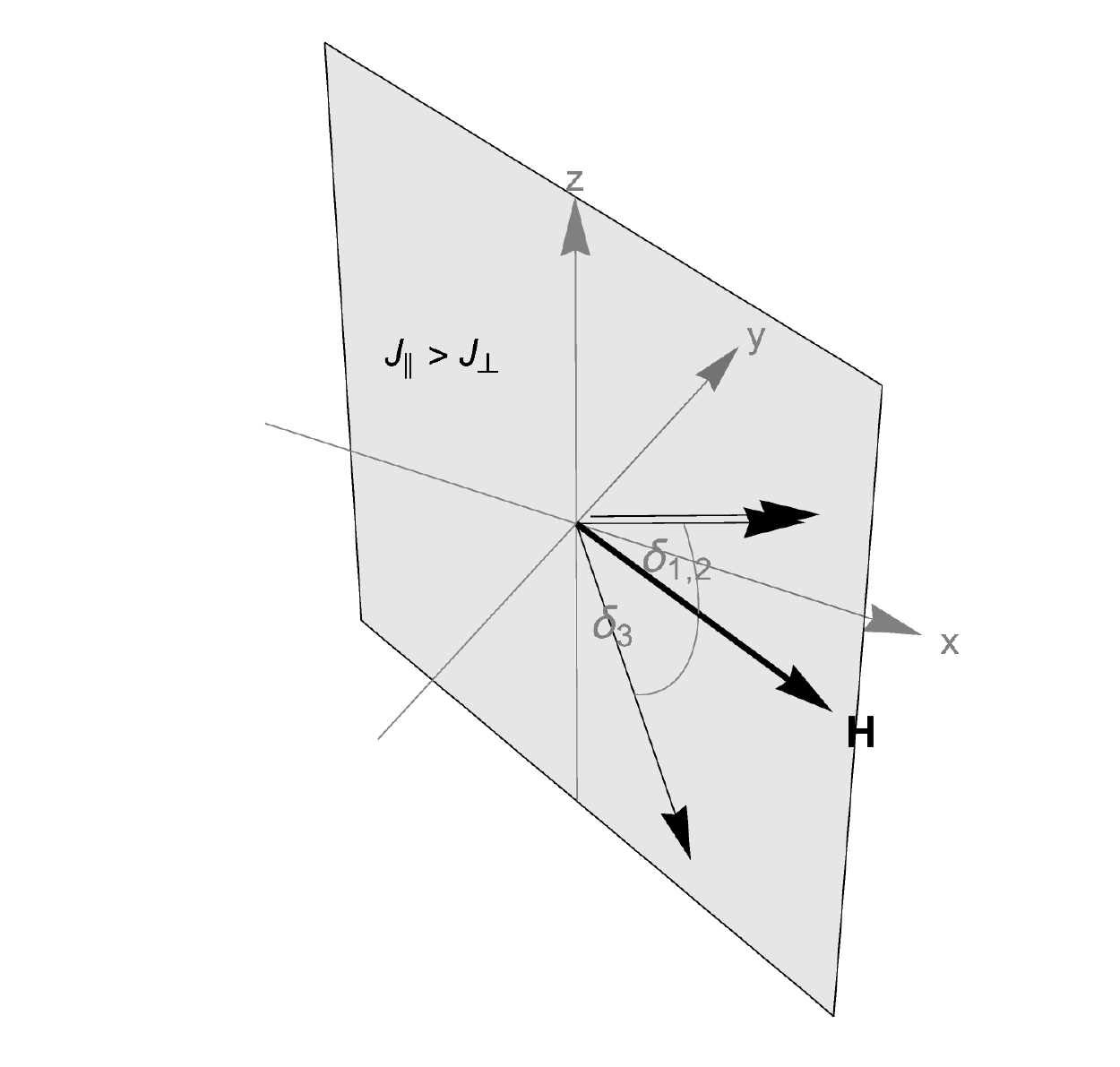}
    \caption{Illustration of the spin configuration near saturation. Solid arrows represent the sublattice moments, the thick arrows the respective magnetic field. Left: field applied parallel to the c direction. The sublattice moments form a cone around the field direction. Middle: field applied perpendicular to the c direction. For $J_\parallel<J_\perp$, the sublattice moments lie in the ab plane indicated by the gray rectangle. Right: field applied perpendicular to the c direction, and $J_\parallel>J_\perp$. The sublattice moments lie in a plane containing the c axis and the field axis, indicated by the gray rectangle.}
    \label{fig:hsat}
\end{figure*}
Table~\ref{tbl:values} also contains the values for the saturation fields $H_\text{sat}^{\parallel,\perp}$ where the field dependent magnetization $M(H)$ per Yb$^{3+}$ ion reaches its saturation values $M_\text{sat}$. For NaYbO$_\text2$ again only powder samples were available. The saturation field $H_\text{sat}^\parallel$ of NaYbS$_\text2$ parallel to the c axis was too high to be reached in our experiments.

$H_\text{sat}$ is defined as an instability of the fully polarized state towards $\Delta m_s=1$ spin flips (magnons). It can be calculated within a classical approximation which is described in Appendix~\ref{sec:hsat}. We parameterize the spins as moment vectors on three interpenetrating sublattices where on each sublattice, the moments are aligned pairwise parallel, and the classical energy density $e={\cal E}/(\nu s)$ is a function of three pairs of polar and azimuthal angles of the sublattice moments.

\subsubsection{Field parallel to the c direction}

A magnetic field applied parallel to the crystallographic c direction preserves the rotational symmetry of the Hamiltonian~(\ref{eqn:cef:ham}). For $s=1/2$, three possible configurations of the sublattice moments near saturation have been found~\cite{yamamoto:14,yamamoto:14-1,yamamoto:17}, called $0$-coplanar ($0<J_\perp/J_\parallel\lesssim3/2$), $\pi$-coplanar ($3/2\lesssim J_\perp/J_\parallel\lesssim2.2$) and umbrella phase ($J_\perp/J_\parallel\gtrsim2.2$) which we assume here. The left sketch in Fig.~\ref{fig:hsat} illustrates this particular configuration. At any finite field value, the sublattice moments arrange on a cone such that the projections onto the ab plane form a 120-degree structure. Polar angles are all equal, $\theta_i=\xi$. Near saturation with $\xi\ll1$, the classical ground-state energy density is given by
\begin{align}
	e_\parallel
	&\approx
	3sJ_\parallel-h_\parallel
	-\xi^2\left(\frac32sJ_\perp+3sJ_\parallel
	-\frac12h_\parallel\right),
	\\
	h_\parallel&=
	g_\parallel\mu_\text B\mu_0H.
	\nonumber
\end{align}
This demonstrates the competition between the best possible antiferromagnetic alignment of the spins with respect to each other and the alignment parallel to the field direction. The first two terms yield the ground-state energy per spin for the fully polarized state: We gain energy $h_\parallel$, but there is an energy loss $3sJ_\parallel$ because of the perfect misalignment (i.\,e. ferromagnetic alignment) of the moments on all three sublattices. The term $\propto\xi^2$ shows what happens when the umbrella opens a bit: We gain energy $(3/2)sJ_\perp\xi^2$ from the (small) 120-degree projections of the spins onto the ab plane perpendicular to the field. This energy gain is half of what we would obtain if it were possible to align the projections in a Néel like manner, impossible on a triangular lattice. We also gain energy $3sJ_\parallel\xi^2$ due to the (small) reduction of the pairwise ferromagnetic spin alignment in the direction parallel to the field on the three sublattices. For the same reason, we lose energy $(1/2)h_\parallel\xi^2$. With the saturation magnetization $M_\text{sat}^{\parallel,\perp}=sg_{\parallel,\perp}\mu_\text B/\text{Yb}^{3+}$ (see Table~\ref{tbl:values} for the experimental values), we obtain
\begin{equation}
    \mu_0H_\text{sat}^\parallel
	=
	\frac{3s^2(2J_\parallel+J_\perp)}{M_\text{sat}^\parallel}.
\end{equation}

\subsubsection{Field in the ab plane}

For $J_\parallel\ne J_\perp$, the magnetic phase diagram has three intermediate phases between $H=0$ and saturation~\cite{kawamura:85,seabra:11,wu:20}. Like in the case $\vec H\parallel\text c$, we assume an umbrella-shape spin structure right below the saturation field, however the umbrella has no rotational symmetry around the axis set by the direction of the magnetic field and might degenerate to a planar configuration~\cite{kawamura:84,rastelli:92,jacobs:93,ohyama:95,nikuni:98,jacobs:02}. In Appendix~\ref{sec:hsat}, we parameterize a spin on sublattice $i$ with its polar and azimuthal angles $\theta_i$ and $\phi_i$. Near saturation, $\theta_i\to\pi/2$, $\phi_i\to\alpha$ where $\alpha$ is the angle of the applied field relative to the a axis, and we obtain
\begin{equation}
    \mu_0H_\text{sat}^\perp
	=
	\frac{9s^2J_\perp}{M_\text{sat}^\perp}
\end{equation}
for the saturation field in the ab plane. This is the case for $J_\parallel<J_\perp$. The independence of $H_\text{sat}^\perp$ from $J_\parallel$ suggests that a planar spin configuration with all spins in the ab plane is energetically favorable at and at least infinitesimally below $H_\text{sat}$. At finite $H<H_\text{sat}$, a distorted cone might form. Writing $\delta_i=\pi/2-\theta_i$, $\epsilon_i=\alpha-\phi_i$, the six angles are a solution to Eq.~(\ref{eqn:cef:mhdet}) with $\delta_i,\epsilon_i\ne0$.

A geometric interpretation can be obtained by looking at two possible planar configurations near $H_\text{sat}$, sketched in the middle and right illustration of Fig.~\ref{fig:hsat}: the sublattice moments form a fan either in the ab plane setting $\delta_i=0$ or in a plane perpendicular to it containing the c and the field axis with $\epsilon_i=0$. For the former, the gradient near $H_\text{sat}$ is given by
\begin{equation}
	\left(\frac{\partial e_\perp}{\partial\epsilon_i}\right)_{\delta_i=0}
	\approx
	\left(\frac{h_\perp}3-2sJ_\perp\right)
	\begin{pmatrix}
		\epsilon_1-u\left(\epsilon_2+\epsilon_3\right)
		\\
		\epsilon_2-u\left(\epsilon_3+\epsilon_1\right)
		\\
		\epsilon_3-u\left(\epsilon_1+\epsilon_2\right)
	\end{pmatrix}
\end{equation}
with $u=sJ_\perp/\left[2sJ_\perp-(1/3)h_\perp\right]$. Minimizing this gives
\begin{equation}
	\epsilon_1
	=
	\epsilon_2
	=
	\frac u{1-u}\epsilon_3,
	\quad
	u
	=
	-1\mbox{ or }\frac12.
\end{equation}
From $u=-1$ we obtain $h_\perp=9sJ_\perp$ (first factor in Eq.~(\ref{eqn:cef:hsatperp})), for $u=1/2$ we obtain the unphysical solution $h_\perp=0$. The ground-state energy density is $e_\perp=3sJ_\perp-h_\perp$. Similar to the case $\vec H\parallel\text c$, we lose energy $3sJ_\perp$ due to the ferromagnetic alignment of all spins, and we gain energy $h_\perp$. Deviating slightly from full polarization, but still with all sublattice moments in the ab plane we gain energy $\Delta e_\perp\propto3sJ_\perp|\epsilon_i|$ due to both the small antiferromagnetic component in the spin alignment and the reduction of the moment parallel to the magnetic field.

The gradient near $H_\text{sat}$ for a spin configuration in the plane containing the c axis and the field ($\epsilon_i=0$) is given by
\begin{equation}
	\left(\frac{\partial e_\perp}{\partial\delta_i}\right)_{\epsilon_i=0}
	\approx
	\left(\frac{h_\perp}3-2sJ_\perp\right)
	\begin{pmatrix}
		\delta_1-u\left(\delta_2+\delta_3\right)
		\\
		\delta_2-u\left(\delta_3+\delta_1\right)
		\\
		\delta_3-u\left(\delta_1+\delta_2\right)
	\end{pmatrix}
\end{equation}
with $u=sJ_\parallel/\left[2sJ_\perp-(1/3)h_\perp\right]$. Minimizing this gives
\begin{equation}
	\delta_1
	=
	\delta_2
	=
	\frac u{1-u}\delta_3,
	\quad
	u
	=
	-1\mbox{ or }\frac12.
\end{equation}
From $u=-1$ we obtain $h_\perp=3s\left(J_\parallel+2J_\perp\right)$ (second factor in Eq.~(\ref{eqn:cef:hsatperp})). This is the case for $J_\parallel>J_\perp$. Deviating from full polarization now results in an energy gain $\Delta e_\perp\propto s(J_\parallel+2J_\perp)|\delta_i|$, which is, for $J_\parallel<J_\perp$, smaller than the energy gain when opening in the ab plane. For $u=1/2$ we obtain $h_\perp=6s\left(J_\perp-J_\parallel\right)$. This solution is unphysical because this would include an energy {\em loss\/} $\propto2sJ_\parallel$ when deviating from full polarization.

In the last six rows of Table~\ref{tbl:calc} we note the values obtained from the equations above. For NaYbS$_2$, $J_\parallel$ cannot be determined because of the missing value for $H_\text{sat}^\parallel$. If we take the exchange constants derived from $I_\text{EPR}(T)$ and $\chi(T)$, we can roughly estimate $56\,\text T\lesssim\mu_0H_\text{sat}^\parallel\lesssim112\,\text T$ which is indeed a larger range than experimentally accessible for us.

\section{Summary and concluding remarks}

In summary, we have shown that the series of NaYbCh$_2$ compounds contain nearly perfect magnetic Yb$^{3+}$ triangular lattice planes. The determination of the crystal field levels by neutron scattering has shown that at temperatures $T\lesssim100\,\rm K$ an application of a pseudospin model is justified.  A consistent treatment in the framework of the anisotropic Heisenberg triangular lattice model with next nearest neighbor (nn) coupling of Yb$^{3+}$ pseudospins of 1/2 gives good agreement with some experimental results. In particular, the estimation of the main exchange energies  (rotationally invariant elements of the exchange matrix) and the derived saturation fields are in good agreement. Nevertheless, the question to the origin of the absence of magnetic order and the emerging spin liquid state is not completely answered. In principle, off diagonal contributions in the exchange matrix can be responsible for it. However, it is not possible to estimate these off diagonal contributions on the basis of the available data. This is not unusual and is also true for the other prominent Yb$^{3+}$ triangular lattice system YbGaMgO$_4$~\cite{li:15,li:19}.

Furthermore, the next-nearest neighbor exchange (nnn) may also play a role. It is already known from the Heisenberg lattice without spin-orbit coupling that the additional frustration introduced with an antiferromagnetic coupling can lead to a quantum spin liquid phase where in most cases already a nnn coupling which is one order of magnitude smaller than the nearest-neighbor coupling is sufficient~\cite{sedrakyan:20}. However, we cannot exclude that the dipolar interaction between the Yb$^{3+}$ ions also plays a role like for example in the YbAlO$_3$ quantum magnet~\cite{li:15}. This long-ranged interaction as an additional competing effect could be another reason for the strong frustration and the suppression of the order.

It must also be noted that the effective exchange constants $J_\parallel$ and $J_\perp$ in the NaYbCh$_2$ materials are comparatively large for an Yb$^{3+}$ triangular lattice. In YbMgGaO$_4$, NaBaYb(BO$_3$)$_2$ and Rb$_3$Yb(PO$_4$)$_2$ the low-temperature Curie-Weiss temperatures and thus the exchange constants are more than one order of magnitude smaller compared to the NaYbCh$_2$ system~\cite{li:15,li:19,guo:19,guo:20}. In this sense, these other systems with small exchange couplings can  be regarded as nearly-single-ion systems (without significant exchange coupling between the ions). This is most evident in the magnetic part of the specific heat $c_\text m(T)$. Here a Schottky peak develops in an applied magnetic field due to the Zeeman splitting of the Yb$^{3+}$ CEF ground-state doublet. This peak shifts with the field to higher temperatures. Also, the saturation field in the magnetization $M(H)$ is in the range of a few Tesla, while in the NaYbCh$_2$ systems the saturation fields exceed $10\,\rm T$.

Furthermore, we have found a pronounced and large linear temperature dependence of the magnetic specific heat $c_\text m(T)=\gamma T$ with a residual value $\gamma\approx1\,{\rm J}/\rm molK^2$ for the NaYbCh$_2$ systems. This is a crucial feature of the (gapless) quantum spin liquid ground state and is consistent with the residual fluctuations detected by muon spin relaxation ($\mu$SR), nuclear magnetic resonance (NMR) and finally inelastic neutron scattering. For the other Yb$^{3+}$ triangular lattice materials mentioned above, such residual contributions are absent or negligibly small. The presence of the large $\gamma$ term is a clear evidence for a gapless spin liquid ground state. In analogy to correlated 4f heavy fermion systems with enhanced renormalized electronic density of states at the Fermi level we have an enhanced (renormalized) density of magnetic \guil{fermion-like} states due to fluctuations associated with the generic spin liquid ground state.

Another difference to other Yb$^{3+}$ triangular lattices is the occurrence of field induced order. This shows that competing interactions are responsible for the spin liquid state and place the systems in the vicinity of a critical point. As already mentioned, this might originate from small but finite symmetry-compatible off-diagonal components in the Yb--Yb exchange matrix and/or a possible next-nearest neighbor interaction. 

Taken together, we have successfully shown that the NaYbCh$_2$ delafossites have their own fascinating physics which differs significantly from the previously known planar Yb$^{3+}$ triangular lattices. There is both exchange anisotropy and spin anisotropy in the systems which is an an essential ingredient and enhancement for frustration and absence of magnetic order. The anisotropy is particularly pronounced in NaYbS$_2$ with a ratio of the coupling coefficients of $J_\perp/J_\parallel=6.25$ and a ratio of the EPR g factors of $g_\perp/g_\parallel=5.6$. We are waiting to see further intriguing developments in the field of the 4f delafossites. For example, the series could be extended to include compounds with the chalcogen Te. These new compounds might exhibit a smaller band gap or even be metallic due to extended Te 5p orbitals. This would establish a bridge between the quantum spin liquid in the Mott insulator and a Fermi liquid in a correlated (semi) metal.  

\begin{acknowledgments}
We are grateful to Sasha Chernychev, Jeff Rau, Oliver Stockert, and Peter Thalmeier for elucidating discussions and valuable comments. This work has been partially funded by the German Research Foundation (DFG) within the Collaborative Research Center SFB 1143 (Projects B03, C02,C03, and A05).
\end{acknowledgments}

\appendix

\section{Stevens operators}
\label{sec:stevens}

The Stevens operators in Eq.~(\ref{eqn:cef:htrig}) have the explicit form~\cite{rotter:17}
\begin{widetext}
\begin{align*}
	O_2^0
	&=
	3J_z^2-j(j+1),
	\\
	O_4^0
	&=
	35J_z^4-30j(j+1)J_z^2+25J_z^2
	-6j(j+1)+3\left[j(j+1)\right]^2,
	\\
	O_4^3
	&=
	\frac14\left[
	J_z\left(J_+^3+J_-^3\right)+\left(J_+^3+J_-^3\right)J_z
	\right],
	\\
	O_6^0
	&=
	231J_z^6-315j(j+1)J_z^4+735J_z^4
	+105\left[j(j+1)\right]^2J_z^2-525j(j+1)J_z^2+294J_z^2
	\nonumber\\
	&\phantom{=}
	-5\left[j(j+1)\right]^3+40\left[j(j+1)\right]^2-60j(j+1),
	\\
	O_6^3
	&=
	\frac14\left\{
	\left[11J_z^3-3j(j+1)J_z-59J_z\right]\left(J_+^3+J_-^3\right)
	+
	\left(J_+^3+J_-^3\right)\left[11J_z^3-3j(j+1)J_z-59J_z\right]
	\right\},
	\\
	O_6^6
	&=
	\frac12\left(J_+^6+J_-^6\right),
\end{align*}
\end{widetext}
and we use the standard definitions
\begin{align*}
	J^2\left|j,m\right\rangle
	&=
	j(j+1)\left|j,m\right\rangle,
	\\
	J_z\left|j,m\right\rangle
	&=
	m\left|j,m\right\rangle,
	\\
	J_+\left|j,m\right\rangle
	&=
	\sqrt{j(j+1)-m(m+1)}\left|j,m+1\right\rangle,
	\\
	J_-\left|j,m\right\rangle
	&=
	\sqrt{j(j+1)-m(m-1)}\left|j,m-1\right\rangle,
	\\
	J_x&=\frac12\left(J_++J_-\right),
	\\
	J_y&=\frac1{2\rm i}\left(J_+-J_-\right).
\end{align*}

\section{Crystal-field Hamiltonian matrices}
\label{sec:ham:mat}

In a cubic environment with a threefold quantization axis, the Hamiltonian matrix derived from Eq.~(\ref{eqn:htri:cubic}) for $j=7/2$ is given by
\begin{widetext}
\begin{align*}
	H_\text{CEF}^{(3)}
	&=
	\left(\begin{smallmatrix}
	420\left(B_4^{(3)}+3B_6^{(3)}\right) & 0 &
		0 & 30\sqrt{70}\left(21B_6^{(3)}-4B_4^{(3)}\right) \\
	0 & -60\left(13B_4^{(3)}+105B_6^{(3)}\right) & 0 & 0\\
	0 & 0 & -180\left(B_4^{(3)}-63B_6^{(3)}\right) & 0\\
	30\sqrt{70}\left(21B_6^{(3)}-4B_4^{(3)}\right) & 0 &
		0 & 180\left(3B_4^{(3)}-35B_6^{(3)}\right)\\
	0 & -15\sqrt{10}\left(16B_4^{(3)}+147B_6^{(3)}\right) & 0 & 0\\
	0 & 0 & 0 & 0\\
	3465\sqrt{7}B_6^{(3)} & 0 &
		0 & 15\sqrt{10}\left(16B_4^{(3)}+147B_6^{(3)}\right)\\
	0 & 3465\sqrt{7}B_6^{(3)} & 0 & 0
   \end{smallmatrix}\right\rfloor
   \nonumber\\
	&\phantom={}
	\left\lceil\begin{smallmatrix}
	0 & 0 & 3465\sqrt{7}B_6^{(3)} & 0 \\
	-15\sqrt{10}\left(16B_4^{(3)}+147B_6^{(3)}\right) & 0 &
		0 & 3465\sqrt{7}B_6^{(3)} \\
	0 & 0 & 0 & 0 \\
	0 & 0 & 15\sqrt{10}\left(16B_4^{(3)}+147B_6^{(3)}\right) & 0 \\
	180\left(3B_4^{(3)}-35B_6^{(3)}\right) & 0 & 
		0 & 30\sqrt{70}\left(4B_4^{(3)}-21B_6^{(3)}\right) \\
	0 & -180\left(B_4^{(3)}-63B_6^{(3)}\right) & 0 & 0 \\
	0 & 0 & -60\left(13B_4^{(3)}+105B_6^{(3)}\right) & 0 \\
	30\sqrt{70}\left(4B_4^{(3)}-21B_6^{(3)}\right) & 0 &
		0 & 420\left(B_4^{(3)}+3B_6^{(3)}\right)
	\end{smallmatrix}\right).
\end{align*}
\end{widetext}
Its eigenvalues are
\begin{align*}
	E_{\Gamma_6}&=-315\left(4B_4^{(3)}+45B_6^{(3)}\right),
	\\
	E_{\Gamma_7}&=\phantom{-}405\left(4B_4^{(3)}-21B_6^{(3)}\right),
	\\
	E_{\Gamma_8}&=-180\left(B_4^{(3)}-63B_6^{(3)}\right),
\end{align*}
the corresponding wavefunctions are given by Eqs.(\ref{eqn:cubicwf}).

In a trigonal CEF, all six crystal-field parameters introduced in Eq.~(\ref{eqn:cef:htrig}) are independent, and we obtain for the $j=7/2$ matrix representation of the Hamiltonian
\begin{widetext}
\begin{align*}
	H_\text{CEF}
	&=
	\left(\begin{smallmatrix}
	21\left(B_2^0+20B_4^0+60B_6^0\right) & 0 &
	0 & \sqrt{35}\left(B_4^3+12B_6^3\right) \\
	0 & 3\left(B_2^0-260B_4^0-2100B_6^0\right) &
	0 & 0 \\
	0 & 0
	& -9\left(B_2^0-30B_4^0+1260B_6^0\right) & 0\\
	6\sqrt{35}\left(B_4^3+12B_6^3\right) & 0 &
	0 & -15\left(B_2^0+36B_4^0-420B_6^0\right) \\
	0 & 12\sqrt{5}\left(B_4^3-21B_6^3\right) &
	0 & 0 \\
	0 & 0 &
	0 & 0 \\
	360\sqrt{7}B_6^6 & 0 &
	0 & 12\sqrt{5}\left(21\sqrt{5}B_6^3-B_4^3\right) \\
	0 & 360\sqrt{7}B_6^6 &
	0 & 0 \\
	\end{smallmatrix}\right\rfloor
	\nonumber\\
	&\phantom={}
	\left\lceil\begin{smallmatrix}
	0 & 0 &
	360\sqrt{7}B_6^6 & 0 \\
	12\sqrt{5}\left(B_4^3-21B_6^3\right) & 0 &
	0 & 360\sqrt{7}B_6^6 \\
	0 & 0 &
	0 & 0 \\
	0 & 0 &
	12\sqrt{5}\left(21\sqrt{5}B_6^3-B_4^3\right) & 0 \\
	-15\left(B_2^0+36B_4^0-420B_6^0\right) & 0 &
	0 & -6\sqrt{35}\left(B_4^3+12B_6^3\right) \\
	0 & -9\left(B_2^0-30B_4^0+1260B_6^0\right) &
	0 & 0 \\
	0 & 0 &
	3\left(B_2^0-260B_4^0-2100B_6^0\right)  & 0 \\
	-6\sqrt{35}\left(B_4^3+12B_6^3\right) & 0 &
	0 & 21\left(B_2^0+20B_4^0+60B_6^0\right) \\
	\end{smallmatrix}\right).
\end{align*}
\end{widetext}
Needless to say that closed-form expressions for the eigenvalues, apart from the pure $\left|7/2,\pm3/2\right\rangle$ doublet, are lengthy and not very insightful.

If we regard the trigonal distortion of the ideal octahedron as small, we obtain corrections to the cubic eigenvalues to first order in the CEF parameters $\delta B_n^m$ like
\begin{widetext}
\begin{align*}
	E_{\Gamma_6}
	&\to E_{\Gamma_6}
	-\frac{70}3\delta\left[14B_4^0-\sqrt2B_4^3
	+10\left(24B_6^0+\sqrt2B_6^3
	+2B_6^6\right)\right],
	\\
	E_{\Gamma_7}
	&\to E_{\Gamma_7}
	+10\delta\left[42B_4^0-3\sqrt2B_4^3
	-14\left(24B_6^0+\sqrt2B_6^3
	+2B_6^6\right)\right],
	\\
	E_{\Gamma_8}
	&\to E_{\Gamma_8}
	\left\{
	\begin{aligned}
	&-9\delta\left[B_2^0+20\left(B_4^0-63B_6^0\right)
	\right]
	\\
	&+9\delta B_2^0+\frac{20}3\delta\left(13B_4^0
	+\sqrt2B_4^3-357B_6^0+56\sqrt2B_6^3
	+112B_6^6\right)
	\end{aligned}\right.,
\end{align*}
\end{widetext}
in particular splitting the $\Gamma_8$ quartet into two Kramers doublets.

\section{Hamiltonian parametrization}
\label{sec:params}

\begin{table}
\caption{Common exchange parametrizations for the triangular-lattice Hamiltonian}
\[
	\begin{array}{l|llll}
		\mbox{Authors} & J_\perp & J_z & J_\Delta & J_{xy} \\
		\hline\hline
		\mbox{Li et al.~\cite{li:15}}
		& 2J_\pm & J_{zz} & 2J_{\pm\pm}	& J_{z\pm}
		\\
		\mbox{Rau and Gingras~\cite{rau:18}}
		& -2J_\pm & J_{zz} & 2J_{\pm\pm} & 2J_{z\pm}
		\\
		\mbox{Zhu et al.~\cite{zhu:18}}
		& J & \Delta J & 4J_{\pm\pm} & 2J_{z\pm}
	\end{array}
\]
\label{tbl:params}
\end{table}
Our definition of the exchange constants occurring in the Hamiltonian~(\ref{eqn:cef:ham}) is made such that it reduces to the standard XXZ model for $J_\Delta=0$ and $J_{yz}=0$ which in turn reduces to the standard isotropic Heisenberg model for $J_\perp=J_z$. Different authors use slightly different parametrizations, shown in Table~\ref{tbl:params} for easy comparison with literature.

\section{Moment calculation}
\label{sec:moment}

In the limit $T\to\infty$, the correlation functions between spins on different sites factorize, only on-site spin-flip correlations retain a finite value. The denominators of Eqs.~(\ref{eqn:cef:mtwo}) and~(\ref{eqn:cef:mfour}) therefore are given by
\begin{align}
	\left\langle
	S_\text{total}^+S_\text{total}^-
	\right\rangle
	&=\sum_{ij}\left\langle S_i^+S_j^-\right\rangle
	\xrightarrow{T\to\infty}\sum_i\left\langle S_i^+S_i^-\right\rangle
	\nonumber\\
	&=\nu\left\langle S_1^+S_1^-\right\rangle
	=\nu\left(\frac12+\left\langle S_i^z\right\rangle\right)
	\to\frac\nu2.
	\nonumber
\end{align}
With a general short-range exchange Hamiltonian we have to evaluate an expression like
\begin{align}
	\MoveEqLeft
	\left\langle
	\left[{\cal H},S_\text{total}^+\right]
	\left[S_\text{total}^-,{\cal H}\right]
	\right\rangle
	\nonumber\\
	&=
	\left\langle\left(
	\sum_{\left\langle ij\right\rangle}\sum_{\alpha\beta}
	J_{\left\langle ij\right\rangle}^{\alpha\beta}
	\sum_k\left[S_i^\alpha S_j^\beta,S_k^+\right]
	\right)
	\right.\nonumber\\&\phantom={}\left.
	\cdot\left(
	\sum_{\left\langle ij\right\rangle}\sum_{\alpha\beta}
	J_{\left\langle ij\right\rangle}^{\alpha\beta}
	\sum_k\left[S_k^-,S_i^\alpha S_j^\beta\right]
	\right)
	\right\rangle
	\label{eqn:cef:m2}
\end{align}
where $J_{\left\langle ij\right\rangle}^{\alpha\beta}$ is the exchange matrix along (not necessarily nearest-neighbor) bond $\left\langle ij\right\rangle$ with spin indices $\alpha\beta$. The inner sum over sites $k$ can be removed applying Jacobi's identity
\[
	\left[S_i^\alpha S_j^\beta,S_k^+\right]
	=
	\left[S_i^\alpha,\left[S_j^\beta,S_k^+\right]\right]
	-\left[S_j^\beta,\left[S_i^\alpha,S_k^+\right]\right]
\]
and the standard spin commutation relations. Together with translational invariance for a Bravais lattice with $\nu$ sites and $z$ neighbors the first term in Eq.~(\ref{eqn:cef:m2}) simplifies to
\begin{align*}
	\MoveEqLeft
	\left[{\cal H},S_\text{total}^+\right]
	\nonumber\\
	&=
	\frac\nu2\sum_{n=2}^{z+1}\sum_{\alpha\beta}
	J_n^{\alpha\beta}
	\left(
	S_1^\alpha\left[S_n^\beta,S_n^+\right]
	+\left[S_1^\alpha,S_1^+\right]S_n^\beta
	\right),
\end{align*}
accordingly for its complex conjugate. The generalization to non-Bravais lattices, not needed in the present context, should be obvious. For the $R\bar3m$ triangular-lattice exchange Hamiltonian with $z=6$ nearest neighbors given by Eq.~(\ref{eqn:cef:ham}), we eventually obtain for $M_2$, Eq.~(\ref{eqn:cef:mtwo}) in the high-temperature limit the expressions~(\ref{eqn:cef:m2para}) and~(\ref{eqn:cef:m2perp}).

Evaluating the commutators in
\begin{align*}
	\left[{\cal H},\left[{\cal H},S_\text{total}^+\right]\right]
	&=
	\sum_{\left\langle mn\right\rangle}\sum_{\gamma\delta}
	J_{\left\langle mn\right\rangle}^{\gamma\delta}
	\sum_{\left\langle ij\right\rangle}\sum_{\alpha\beta}
	J_{\left\langle ij\right\rangle}^{\alpha\beta}
	\nonumber\\
	&\phantom={}
	\left(
	S_m^\gamma\left[
	S_n^\delta,S_i^\alpha\left[S_j^\beta,S_j^+\right]
	\right]
	\right.
	\nonumber\\
	&\phantom=\left.{}
	+\left[S_m^\gamma,S_i^\alpha\left[S_j^\beta,S_j^+\right]
	\right]S_n^\delta
	\right.
	\nonumber\\
	&\phantom=\left.{}
	+S_m^\gamma\left[
	S_n^\delta,\left[S_i^\alpha,S_i^+\right]S_j^\beta\right]
	\right.
	\nonumber\\
	&\phantom=\left.{}
	+\left[S_m^\gamma,\left[S_i^\alpha,S_i^+\right]S_j^\beta
	\right]S_n^\delta
	\right)
\end{align*}
and its complex conjugate we also use Jacobi's identity, the spin commutation relations and translational invariance. With
\begin{align*}
	T_1
	&=
	\sum_{\left\langle mn\right\rangle}\sum_{\gamma\delta}
	J_{\left\langle mn\right\rangle}^{\gamma\delta}
	\sum_{\left\langle ij\right\rangle}\sum_{\alpha\beta}
	J_{\left\langle ij\right\rangle}^{\alpha\beta}
	S_m^\gamma
	\delta_{ni}\left[S_i^\delta,S_i^\alpha\right]
	\left[S_j^\beta,S_j^+\right]
	\\
	&=\frac \nu4
	\sum_{\delta m=2}^{z+1}\sum_{\gamma\delta}
	J_{\delta m}^{\gamma\delta}
	\sum_{\delta j=2}^{z+1}\sum_{\alpha\beta}
	J_{\delta j}^{\alpha\beta}
	S_{\delta m}^\gamma\left[S_1^\delta,S_1^\alpha\right]
	\left[S_{\delta j}^\beta,S_{\delta j}^+\right],
	\\
	T_2
	&=
	\sum_{\left\langle mn\right\rangle}\sum_{\gamma\delta}
	J_{\left\langle mn\right\rangle}^{\gamma\delta}
	\sum_{\left\langle ij\right\rangle}\sum_{\alpha\beta}
	J_{\left\langle ij\right\rangle}^{\alpha\beta}
	S_m^\gamma\delta_{nj}S_i^\alpha\left[S_j^\delta,
	\left[S_j^\beta,S_j^+\right]\right]
	\\
	&=\frac \nu4
	\sum_{\delta m=2}^{z+1}\sum_{\gamma\delta}
	J_{\delta m}^{\gamma\delta}
	\sum_{\delta i=2}^{z+1}\sum_{\alpha\beta}
	J_{\delta i}^{\alpha\beta}
	S_{\delta m}^\gamma S_{\delta i}^\alpha
	\left[S_1^\delta,\left[S_1^\beta,S_1^+\right]\right],
	\\
	T_3
	&=
	\sum_{\left\langle mn\right\rangle}\sum_{\gamma\delta}
	J_{\left\langle mn\right\rangle}^{\gamma\delta}
	\sum_{\left\langle ij\right\rangle}\sum_{\alpha\beta}
	J_{\left\langle ij\right\rangle}^{\alpha\beta}
	\delta_{mi}
	\left[S_i^\gamma,S_i^\alpha\right]
	\left[S_j^\beta,S_j^+\right]S_n^\delta
	\\
	&=\frac \nu4
	\sum_{\delta n=2}^{z+1}\sum_{\gamma\delta}
	J_{\delta n}^{\gamma\delta}
	\sum_{\delta j=2}^{z+1}\sum_{\alpha\beta}
	J_{\delta j}^{\alpha\beta}
	\left[S_1^\gamma,S_1^\alpha\right]
	\left[S_{\delta j}^\beta,S_{\delta j}^+\right]
	S_{\delta n}^\delta,
	\\
	T_4
	&=
	\sum_{\left\langle mn\right\rangle}\sum_{\gamma\delta}
	J_{\left\langle mn\right\rangle}^{\gamma\delta}
	\sum_{\left\langle ij\right\rangle}\sum_{\alpha\beta}
	J_{\left\langle ij\right\rangle}^{\alpha\beta}
	\delta_{mj}
	S_i^\alpha\left[S_j^\gamma,
	\left[S_j^\beta,S_j^+\right]\right]S_n^\delta
	\\
	&=\frac \nu4
	\sum_{\delta n=2}^{z+1}\sum_{\gamma\delta}
	J_{\delta n}^{\gamma\delta}
	\sum_{\delta i=2}^{z+1}\sum_{\alpha\beta}
	J_{\delta i}^{\alpha\beta}
	S_{\delta i}^\alpha\left[S_1^\gamma,
	\left[S_1^\beta,S_1^+\right]\right]S_{\delta n}^\delta,
	\\
	T_5
	&=
	\sum_{\left\langle mn\right\rangle}\sum_{\gamma\delta}
	J_{\left\langle mn\right\rangle}^{\gamma\delta}
	\sum_{\left\langle ij\right\rangle}\sum_{\alpha\beta}
	J_{\left\langle ij\right\rangle}^{\alpha\beta}
	S_m^\gamma\delta_{ni}
	\left[S_i^\delta,\left[S_i^\alpha,S_i^+\right]\right]
	S_j^\beta
	\\
	&=\left(T_2\right)_{i\leftrightarrow j}^{\alpha
	\leftrightarrow\beta},
	\\
	T_6
	&=
	\sum_{\left\langle mn\right\rangle}\sum_{\gamma\delta}
	J_{\left\langle mn\right\rangle}^{\gamma\delta}
	\sum_{\left\langle ij\right\rangle}\sum_{\alpha\beta}
	J_{\left\langle ij\right\rangle}^{\alpha\beta}
	S_m^\gamma\delta_{nj}
	\left[S_i^\alpha,S_i^+\right]
	\left[S_j^\delta,S_j^\beta\right]
	\\
	&=\left(T_1\right)_{i\leftrightarrow j}^{\alpha
	\leftrightarrow\beta},
	\\
	T_7
	&=
	\sum_{\left\langle mn\right\rangle}\sum_{\gamma\delta}
	J_{\left\langle mn\right\rangle}^{\gamma\delta}
	\sum_{\left\langle ij\right\rangle}\sum_{\alpha\beta}
	J_{\left\langle ij\right\rangle}^{\alpha\beta}
	\delta_{mi}
	\left[S_i^\gamma,\left[S_i^\alpha,S_i^+\right]\right]
	S_j^\beta S_n^\delta
	\\
	&=\left(T_4\right)_{i\leftrightarrow j}^{\alpha
	\leftrightarrow\beta},
	\\
	T_8
	&=
	\sum_{\left\langle mn\right\rangle}\sum_{\gamma\delta}
	J_{\left\langle mn\right\rangle}^{\gamma\delta}
	\sum_{\left\langle ij\right\rangle}\sum_{\alpha\beta}
	J_{\left\langle ij\right\rangle}^{\alpha\beta}
	\delta_{mj}
	\left[S_i^\alpha,S_i^+\right]
	\left[S_j^\gamma,S_j^\beta\right]S_n^\delta
	\\
	&=\left(T_3\right)_{i\leftrightarrow j}^{\alpha
	\leftrightarrow\beta}
\end{align*}
the commutator may be written as
\[
	\left[{\cal H},\left[{\cal H},S_\text{total}^+\right]\right]
	=
	\sum_{\ell=1}^8T_\ell=2\sum_{\ell=1}^4T_\ell,
\]
where the last equality holds for inversion-symmetric exchange only (which is the case here). Then
\[
	\left\langle
	\left[{\cal H},\left[{\cal H},S_\text{total}^+\right]\right]
	\left[\left[S_\text{total}^-,{\cal H}\right],{\cal H}\right]
	\right\rangle
=
	4\sum_{\ell=1}^4\sum_{\ell'=1}^4\left\langle
	T_\ell T_{\ell'}^\dagger
	\right\rangle
\]
has to be evaluated. The final results for the expectation value for $M_4$ in the limit $T\to\infty$ are reproduced in Eqs.~(\ref{eqn:cef:m4para}) and~(\ref{eqn:cef:m4perp}).

\section{Saturation field}
\label{sec:hsat}

To calculate the saturation field of the magnetization, we regard the pseudospins of the triangular lattice as classical vectors living on a three-sublattice structure. On each sublattice $i$, any two spins are aligned parallel relative to each other. The direction of each spin defines the $z'$ axis of a local coordinate system, characterized by polar and azimuthal angels $\Theta_i$ and $\phi_i$. The full Hamiltonian with this parameterization
\begin{widetext}
\[
    \left(
    \begin{array}{c}
  S_{i}^{x}\\
  S_{i}^{y}\\
  S_{i}^{z}
    \end{array}
    \right) =
    \left(
    \begin{array}{ccc}
  \cos\phi_i & -\sin\phi_i & 0\\
  \sin\phi_i & \cos\phi_i & 0\\
  0 & 0 & 1
    \end{array}
    \right)
    \left(
    \begin{array}{ccc}
  \cos\theta_i & 0 & \sin\theta_i\\
  0 & 1 & 0\\
  -\sin\theta_i & 0 & \cos\theta_i
    \end{array}
    \right)
    \left(
    \begin{array}{c}
  S_{i}^{x'}\\
  S_{i}^{y'}\\
  S_{i}^{z'}
    \end{array}
    \right)
\]
\end{widetext}
and $\vec S'=(0,0,s)^T$ is given below.

\subsection{Field parallel to the crystallographic c axis}

With a magnetic field in $z$ direction, the classical energy density per spin is
\begin{widetext}
\begin{align}
		e_\parallel
		=
		\frac{{\cal E}_{z}}{\nu s}
		&=
		sJ_\perp\left[
		\sin\theta_1\sin\theta_2
		\cos\left(\phi_1-\phi_2\right)
		+\sin\theta_2\sin\theta_3
		\cos\left(\phi_2-\phi_3\right)
		+\sin\theta_3\sin\theta_1
		\cos\left(\phi_3-\phi_1\right)
		\right]
		\nonumber\\
		&
		+sJ_\parallel\left[
		\cos\theta_1\cos\theta_2
		+\cos\theta_2\cos\theta_3
		+\cos\theta_3\cos\theta_1
		\right]
		-\frac13h_\parallel\left[
		\cos\theta_1+\cos\theta_2
		+\cos\theta_3\right],
		\label{eqn:cef:epar}
		\\
		h_\parallel
		&=g_\parallel\mu_\text B\mu_0H,
		\nonumber
\end{align}
\end{widetext}
independent of $J_\Delta$ and $J_{yz}$: We assume a three-sublattice structure and positive gyromagnetic ratios, the direction-dependent terms $\propto J_\Delta, J_{yz}$ cancel exactly upon summation over the three sublattice pairs $(1,2)$, $(2,3)$, and $(3,1)$.  According to Eq.~(\ref{eqn:cef:ham}), $J_\Delta$ and $J_{yz}$ give the energy gain when coupling states differing by $\Delta S_z=1$ or $\Delta S_z=2$. For the classical model, the total spin $S_z^\text{tot}=\sum_{i=1}^\nu S_i^z$ is conserved, therefore $J_\Delta$ and $J_{yz}$ cannot contribute to the ground-state energy. The energy is symmetric with respect to the exchange of any two sublattice labels $i=1,2,3$, so the saturation field (as any other ground-state property) cannot depend on it.

The saturation field is given by a singularity of the Hessian matrix
\begin{align}
	m_\text H(e)
	&=\frac{\partial^2e}{\partial\{\delta_i,\phi_i\}
	\partial\{\delta_j,\phi_j\}},
	\nonumber\\
	\det m_\text H(e)
	&=
	0.
	\label{eqn:cef:mhdet}
\end{align}
By symmetry, for the classical ground state of Eq.~(\ref{eqn:cef:epar}), polar angles are all $\theta_i=x$, azimuthal angles are $\alpha$ and $\alpha\pm2\pi/3$ such that $|\phi_i-\phi_j|=2\pi/3$. The nonzero \guil{block} of the Hessian matrix $m_\text H$ reduces to a scalar, $m_\text H(e_\parallel)=\partial^2e_\parallel/\partial x^2$, and Eq.~(\ref{eqn:cef:mhdet}) reduces to $(3/2)sJ_\perp+3sJ_\parallel-(1/2)h_\parallel=0$ which gives
\begin{equation}
	\mu_0H_\text{sat}^\parallel
	=
	\frac{3s}{g_\parallel\mu_\text B}\left(J_\perp+2J_\parallel\right).
	\label{eqn:cef:hsatz}
\end{equation}

\subsection{Field in the ab plane}

Assume the magnetic field $\vec H$ lies in the ab plane at an angle $\alpha$ relative to the $x$ direction. The classical energy density per spin then has the form
\begin{widetext}
\begin{align}
		\frac{{\cal E}_\perp}{\nu s} & =
		sJ_\perp\left[
		\sin\theta_1\sin\theta_2
		\cos\left(\phi_1-\phi_2\right)
		+\sin\theta_2\sin\theta_3
		\cos\left(\phi_2-\phi_3\right)
		+\sin\theta_3\sin\theta_1
		\cos\left(\phi_3-\phi_1\right)
		\right]
		\nonumber\\
		&
		+sJ_\parallel\left[
		\cos\theta_1\cos\theta_2
		+\cos\theta_2\cos\theta_3
		+\cos\theta_3\cos\theta_1
		\right]
		\nonumber\\
		&
		-\frac13h_\perp\left[
		\sin\theta_1\cos\left(\alpha-\phi_1\right)
		+\sin\theta_2\cos\left(\alpha-\phi_2\right)
		+\sin\theta_3\cos\left(\alpha-\phi_3\right)
		\right],
		\\
		h_\perp
		&=
		g_\perp\mu_\text B\mu_0H,
\end{align}
\end{widetext}
independent of $J_\Delta$ and $J_{yz}$ as well. We may assume an umbrella-shape spin structure right below the saturation field, however for $J_\parallel\ne J_\perp$ the umbrella cannot have rotational symmetry around the axis set by the direction of the magnetic field, and according to work done in the context of (Cs,Rb)CuCl$_3$ on similar Heisenberg models energetically favorable spin configurations at high fields are planar~\cite{kawamura:84,rastelli:92,jacobs:93,ohyama:95,nikuni:98,jacobs:02}. Therefore we don't make assumptions about relations between the six angles $\{\theta_i,\phi_i\}$. With $\delta_i=\pi/2-\theta_i$, $\epsilon_i=\alpha-\phi_i$ we write
\begin{align}
		e_\perp
		&=
		\frac{{\cal E}_\perp}{\nu S}
		=
		e_{12}+e_{23}+e_{31},
		\\
		e_{ij}
		&=
		sJ_\perp\cos\delta_i\cos\delta_j
		\cos\left(\epsilon_i-\epsilon_j\right)
		+sJ_\parallel\sin\delta_i\sin\delta_j
		\nonumber\\
		&\phantom{=}
		-\frac16h_\perp\left(
		\cos\delta_i\cos\epsilon_i
		+\cos\delta_j\cos\epsilon_j\right)
		\\
		&=
		e_{ji}.
		\nonumber
\end{align}
In the limit $\delta_i,\epsilon_i\to0$ (full polarization), Eq.~(\ref{eqn:cef:mhdet}) for $m_\text H(e_\perp)$ reads
\begin{align}
	\left(h_\perp-9sJ_\perp\right)^2
	\left[h_\perp-3s\left(J_\parallel+2J_\perp\right)\right]^2&\times
	\nonumber\\
	\left[h_\perp-6s\left(J_\perp-J_\parallel\right)\right]
	h_\perp
	&=0.
	\label{eqn:cef:hsatperp}
\end{align}
In our case, we have $J_\perp>J_\parallel>0$, and the first factor determines the saturation field when lowering the field from values above saturation. We obtain 
\begin{equation}
	\mu_0H_\text{sat}^\perp
	=
	\frac{9sJ_\perp}{\mu_\text Bg_\perp}.
	\label{eqn:cef:hsatxy}
\end{equation}

\bibliography{references}

\end{document}